\documentclass[useAMS,usenatbib]{mn2e}

\usepackage[dvips]{graphicx}
\usepackage{amssymb}

\def\Teff{\ifmmode{T_{\mathrm{eff}}}\else$T_{\mathrm{eff}}$\fi}

\title[Fundamental parameters of M dwarfs]{M dwarfs: effective
  temperatures, radii and metallicities}
\author[Casagrande, Flynn, Bessell]{Luca Casagrande$^{1}$, 
Chris Flynn$^{1}$, Michael Bessell$^{2}$\\
$^1$ Tuorla Observatory, Department of Physics and Astronomy, University of 
Turku, V\"ais\"al\"antie 20, FI-21500, Piikki\"o, Finland\\
$^2$ Research School of Astronomy \& Astrophysics, 
Australian National University, Cotter Road, Weston, ACT 2611, Australia}

\begin{document}

\maketitle

\begin{abstract}

We empirically determine effective temperatures and bolometric luminosities
for a large sample of nearby M dwarfs, for which high accuracy optical and
infrared photometry is available. We introduce a new technique which exploits
the flux ratio in different bands as a proxy of both effective temperature and
metallicity. Our temperature scale for late type dwarfs extends well below 
$3000\,\rm{K}$ (almost to the brown dwarf limit) and is supported by 
interferometric angular
diameter measurements above $3000\,\rm{K}$. Our metallicities are in excellent
agreement (usually within 0.2~dex) with recent determinations via 
independent techniques. A subsample of cool M dwarfs with metallicity 
estimates based on hotter \emph{Hipparcos} common proper--motion companions 
indicates our metallicities are also reliable below $3000\,\rm{K}$, a 
temperature range unexplored
until now. The high quality of our data allow us to identify a striking feature
in the bolometric luminosity versus temperature plane, around the transition
from K to M dwarfs. We have compared our sample of stars with theoretical
models and conclude that this transition is due to an increase in the radii of 
the M dwarfs, a feature which is not reproduced by theoretical models.

\end{abstract}

\begin{keywords}

\end{keywords}

\section{Introduction}

Low--mass dwarfs are the dominant stellar component of the Galaxy and have been
employed in a variety of Galactic studies: tracing Galactic disk kinematics
(e.g. Hawley, Gizis \& Reid 1996; 
Gizis et al. 2002; Bochanski et al. 2005, 2007),
studying the stellar age--velocity relations (West et al. 2006), investigating
Galactic structure (e.g. Reid et al. 1997, Kerber et al. 2001, Pirzkal et
al. 2005), and the Galaxy's mass and luminosity (e.g. Hawkins \& Bessell 1988;
Kirkpatrick et al. 1994; Gould, Bahcall \& Flynn 1997, 1998; Zheng et al. 2001,
2004). An increasing number of M dwarfs are now known to host exoplanets
(e.g. Delfosse et al. 1998; Butler et al. 2004; Bonfils et al. 2007,
Udry et al. 2007). The determination of accurate fundamental parameters for M
dwarfs has therefore relevant implications for both stellar and Galactic
astronomy.

Observationally, the spectra of these stars are marked by the presence of
strong molecular absorption features, in either optical (e.g. TiO and VO) and
infrared regions (e.g. $\textrm{H}_2\textrm{O}$ and CO). 
Molecular lines blend with
all other lines and create a pseudo--continuum, rendering all spectral analysis
difficult (e.g.\ Gustafsson 1989). 
However, recent advances in model atmospheres of low--mass dwarf
stars (Hauschildt et al. 2003, Brott \& Hauschildt 2005), have boosted the
number of studies deriving accurate metallicities for M dwarfs (Woolf \&
Wallerstein 2005, 2006; Bean et al. 2006a, 2006b). Modeling the internal
structure, atmospheric properties and magnetic activity of M dwarfs
(e.g. Burrows et al. 1993, Allard et al. 1997, Baraffe et al. 1998, Hauschildt,
Allard \& Baron 1999, Allred et al. 2006, Reiners \& Basri 2007) 
also presents ongoing theoretical challenges.
For a small number of nearby M dwarfs, interferometry is currently providing
direct angular diameter measurements (Lane, Boden \& Kulkarni 2001; S\'egransan
et al. 2003; Berger et al. 2006) which confirm a large discrepancy between the
predicted and observed radii, as has been noted in eclipsing binaries with
M--type components (see Ribas 2006 for a review).

In this paper, we empirically determine the effective temperatures and the
bolometric luminosities for more than 340 M dwarfs. This work is an extension
of our previous study on G and K dwarfs, to which we applied the InfraRed Flux
Method (IRFM, Casagrande, Portinari \& Flynn 2006).  The effective temperature
and the bolometric luminosity scales we derive are accurate at a level of a few
percent and are supported by interferometric angular diameters above 
$\sim 3000\,\rm{K}$.

We find in this study that below about $4000\,\rm{K}$ the monochromatic to 
bolometric flux ratio in different bands is a proxy of both effective
temperature and metallicity. By exploiting this feature, we are able to derive
not only $\Teff$, but also the metallicities of our 
M dwarfs, which are found to be in very good agreement (usually within 
$0.2-0.3$~dex)
with those inferred using the Bonfils et al. (2005) calibration or directly
measured from Woolf \& Wallerstein (2005, 2006). The technique we propose 
also looks promising for stars much cooler than those explored in the 
aforementioned studies.

We find considerable structure in the temperature--luminosity plane, especially
around the transition between K and M dwarfs. Our study circumstantially
confirms previous works which indicate that the radii of M dwarfs are larger by
about 15\% than model predictions. We also find strong evidence that 
this discrepancy, clearly observed in M dwarfs in eclipsing binary systems, 
is also present in nearby disk M dwarfs.

The paper is organized as follows. In Section \ref{sample} we describe our
sample of M dwarfs and in Section \ref{phoenix} we compare it with the Phoenix
model atmosphere in the two--colour plane. We then review the IRFM and present
our new technique for estimating effective temperatures and metallicities below
$4000\,\rm{K}$ in Section \ref{mfm}. Our proposed metallicity, bolometric 
luminosity and effective temperature scales along with the comparison with 
other empirical determinations are discussed in Section \ref{abs_cor}, 
\ref{luca} and \ref{tesca}, respectively. In
Section \ref{jump} we analyze the stars with good \emph{Hipparcos} parallaxes
in the HR diagram. We find a strong feature which marks the transition from K
to M dwarfs and which is due to an increase in the observed stellar radii. 
We briefly 
discuss possible reasons, including the effect of magnetic fields and 
molecular opacity. We finally conclude in Section \ref{conclu}.

\section{The sample}\label{sample}

Our basic sample consists of 343 nearby M dwarfs with high quality optical and 
infrared photometry. We describe the sample selection in this Section.

\subsection{Johnson--Cousins photometry}\label{jc}

In recent years major efforts have been devoted to obtaining high accuracy
Johnson--Cousins photometry for cool stars (Kilkenny et al. 1998, 2007; Koen et
al. 2002). One third of our sample is built from the extensive work of Koen et
al. (2002) who presented homogeneous and standardized $UBV(RI)_C$ photometry
for more than 500 M stars, half of which are main sequence dwarfs.  Variability
has a very high incidence among cool stars; however the existence of $\sim 100$
or more \emph{Hipparcos} measurements per star spread over several years,
together with the excellent temporal stability of the magnitude scale, permits
the detection of variability at the level of few hundredths of a magnitude. The
red standards provided by Koen et al. (2002) are all non-variables in this
sense (i.e. the \emph{Hipparcos} variability flag is not set); Koen et
al. (2002) also provide the SIMBAD and CCDM (Dommanget \& Nys 1994)
classification for variability and double/multiple stars. We have discarded all
star having those labels, for a total final sample 128 M dwarfs. Besides
Johnson--Cousins photometry accurate to 0.01~mag or better, all these stars 
have \emph{Hipparcos} parallaxes better than 15\% and are all closer than 60 pc.

Another accurate source of Johnson--Cousins photometry for M dwarfs is Reid et
al. (2002, 2003, 2004). Altogether, they provided new $V(RI)_C$ (and $B$ for
the brightest sources) measurements for more than 370 NLTT stars.  We have used
the SIMBAD classification plus the essential notes as given in the
aforementioned papers to remove double, variables, flares, possible
misclassification and stars with nearby companions (or background stars) that
could affect the photometric measurements.  The photometry is accurate and
consistent with the standard photometric system to better than 1\% (Reid et
al. 2002) and with a typical uncertainty less than 2\% even for single night
observations (Reid et al. 2003, 2004). Furthermore, all observations are done
with the same instrument and reduced using identical methods, similar to those
described in Kilkenny et al. (1998) and Koen et al. (2002).  Reid et al. (2002,
2003, 2004) also provide photometric distances for all the stars: although such
estimates do not provide definitive distance measurements, we have used them to
ensure that all the stars chosen are closer than 50 pc and their absolute
magnitudes are consistent with those expected from dwarf stars. Altogether, we
have retained 94 stars with $BV(RI)_C$ and 157 with $V(RI)_C$ photometry.

We also took a few very red dwarfs from Henry et al. (2004) who measured
$V(RI)_C$ colours using standards from Graham (1982), Bessell (1990a) and
Landolt (1992). These stars are also generally fainter and the accuracy is
somewhat lower, with a typical uncertainty of $\pm 0.03$ mag in each band
(Henry et al. 2004). We have selected 11 such stars, all closer than 40
pc. Finally, another source of $V(RI)_C$ photometry for very red stars is from
Bessell (1991), from which we took 5 stars. For the stars in Henry et
al. (2004) and Bessell (1991) we have used the SIMBAD classification to avoid
flares or variables, although in this case the most affected bands would 
be the blue ones ($U$ and $B$) which we do not have for these stars. 

\begin{figure*}
\begin{center}
\includegraphics[scale=0.6]{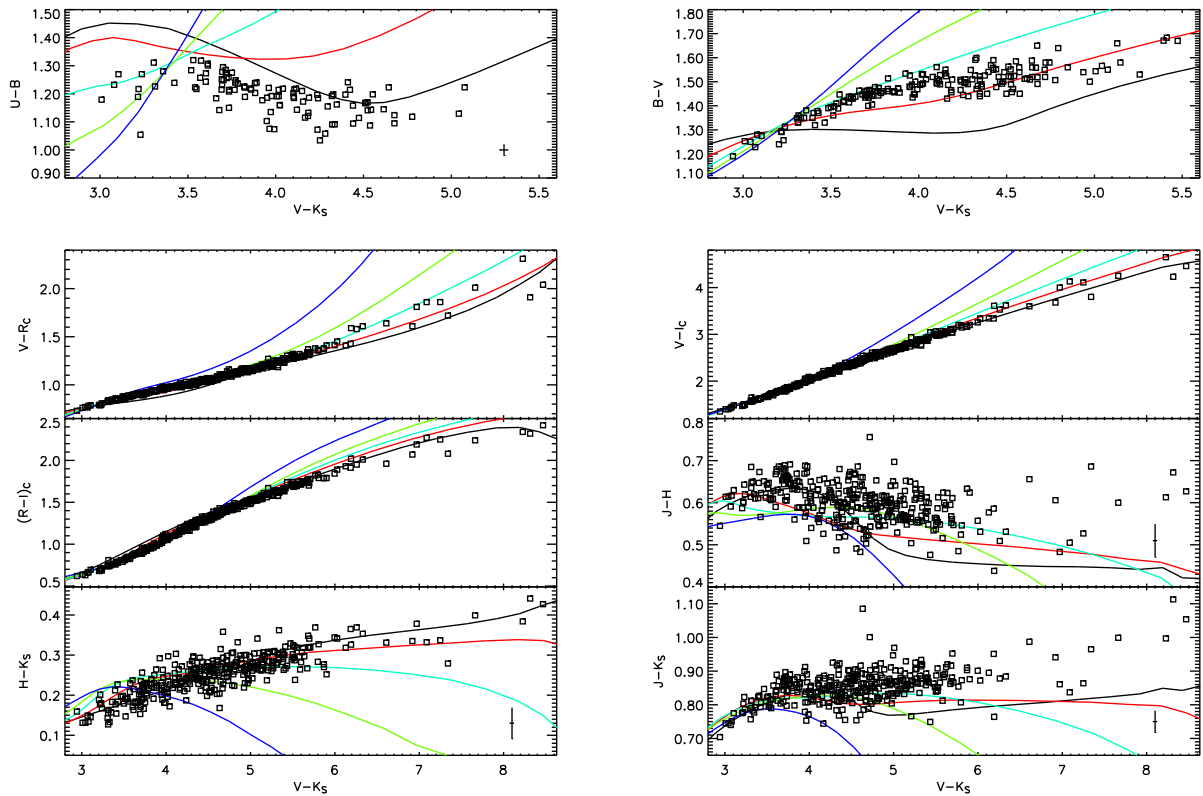}
\caption{Synthetic optical and IR colours from the Phoenix models compared to
our 343 sample stars of Section \ref{sample}. Lines correspond to
models with [M/H] equal $+0.5$~(black), $+0.0$~(red), $-0.5$~(cyan),
$-1.0$~(green) and $-1.5$~(blue). Since metallicities for M dwarfs are usually
not available or are very uncertain, we have not used any metallicity bin for
the stars (open squares). Typical error bars are shown only for certain
indices; in the other bands error bars are comparable or smaller than the size
of the plotting symbols, with very red stars (from Henry et al. 2004) which
are somewhat less accurate.}
\label{vkplot}
\end{center}
\end{figure*}

\subsection{Near--Infrared photometry}\label{nir}

All the stars presented in Section \ref{jc} have $JHK_S$ photometry from
2MASS. In what follows we will use only stars with total photometric errors 
(as given from 2MASS) smaller than 0.10 mag 
(i.e. ``j\_''$+$``h\_''$+$``k\_msigcom''$<0.10$), thus reducing our final 
sample to 343 stars. The typical errors in
$J$ and $K_S$ are around 0.023 mag, whereas a slightly larger uncertainty
(0.032 mag) affects $H$ band photometry.

For most of the Koen et al. (2002) stars, excellent SAAO $JHK$ photometry is
also available (Kilkenny et al. 2007) : this additional photometry is very
valuable to check the accuracy of the 2MASS zero-points and the dependence of
the proposed temperature scale on the adopted absolute calibration. As we
show in Appendix B, the SAAO IR photometry confirms the adopted 2MASS
zero-points and absolute calibration.

\subsection{Metallicities}\label{met_bon}

Measuring metallicities for M dwarfs is still challenging. With decreasing
temperature, the spectra show increasingly abundant diatomic and triatomic
molecules. The molecular bands complicate the calculation of stellar model
atmospheres and cause line blends, making it difficult to estimate the true
continuum and to measure atomic line-strengths over large regions of the
visible spectra.

Major advances in the field have been recently obtained using high--resolution
spectra to measure equivalent widths of atomic lines in regions not dominated
by molecular bands (Woolf \& Wallerstein 2005, 2006). Bonfils et al. (2005)
have measured the metallicity in 20 binary systems, having an M dwarf secondary
and a warmer primary (for which a metallicity is much more readily
obtained). They combined their results with the abundances measured from Woolf
\& Wallerstein (2005) to calibrate an absolute $K_S$ band luminosity versus
colour relation ($M_{K_S}$ and $(V-K_S)$) as a function of metallicity. This
results in a metallicity relation for M dwarfs, but since it depends on
absolute magnitude, we apply it only to those of our stars with accurate
parallaxes available from \emph{Hipparcos}. Altogether in our sample there are
118 stars with accurate 2MASS photometry (see Section \ref{nir}), 
\emph{Hipparcos} parallaxes better
then 15\% and within the range of applicability of the Bonfils et al. (2005)
calibration. The formal accuracy of the relation is $\pm 0.2$~dex, however the
uncertainty in parallaxes introduces an additional error that in the worst case
can be of the same magnitude. Therefore, we expect these metallicities to be
accurate to $0.2-0.3$~dex.

In Section \ref{mfm} we will use these metallicities to calibrate our method to
estimate [M/H] for the rest of the M dwarfs\footnote{The calibration of Bonfils
et al. (2005) returns [Fe/H], whereas the model atmospheres are given for the
total heavy elements fraction [M/H]. For low values of alpha--enhancement, the
difference between the two is negligible, particularly since metallicity
measurements in M dwarfs are still uncertain. In the rest of the paper we will
refer to [M/H], although this is in practice [Fe/H] when we refer to empirical
measurements.}.

\begin{figure*}
\begin{center}
\includegraphics[scale=0.6]{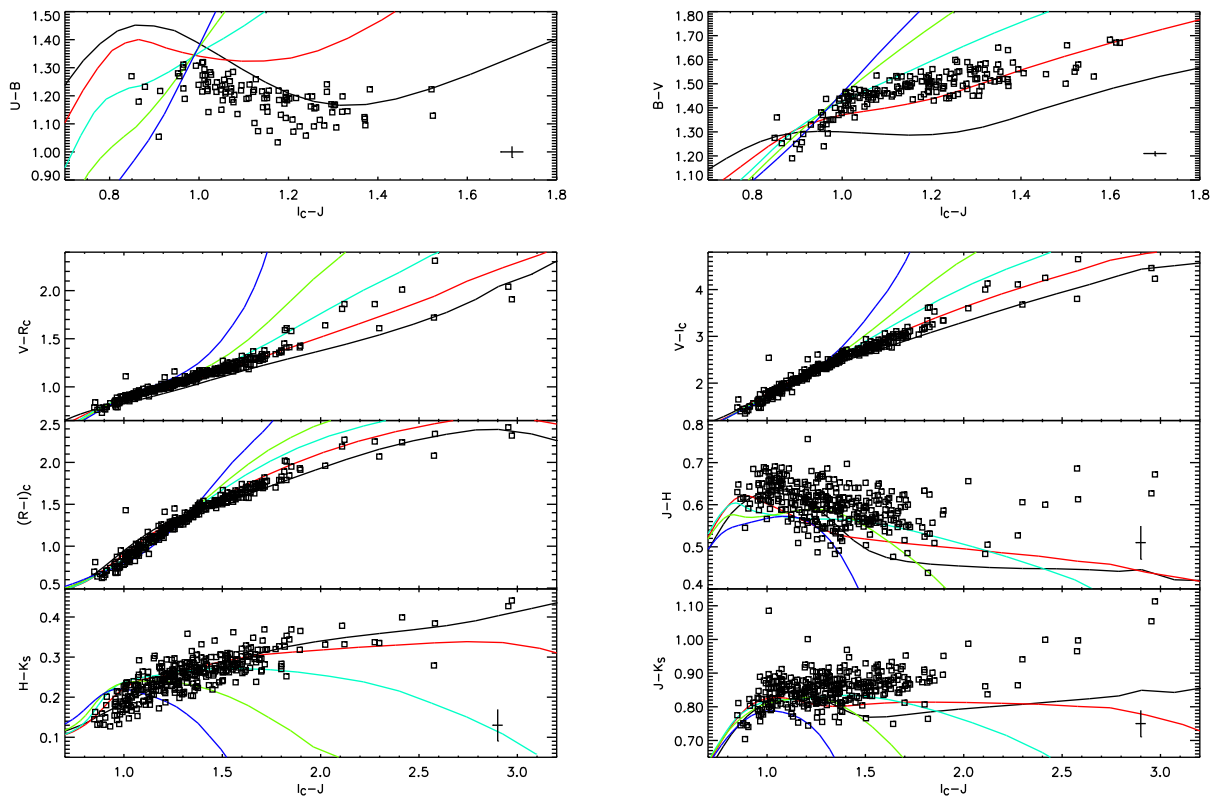}
\caption{Same as Figure \ref{vkplot} but as function of $I_C-J$ colour index.}
\label{ijplot}
\end{center}
\end{figure*}

\section{The Phoenix model atmospheres}\label{phoenix}

We apply an extension of the IRFM we developed for G and K dwarfs (Casagrande
et al. 2006) to M dwarfs. While most of the bolometric flux of the stars is
emitted in the optical and infrared and is covered by our observations, as in
the earlier work, we use model atmospheres to estimate the small part
(typically less than 20\%) of the flux which is outside our observational
bands.

Pioneering work on M dwarfs model atmospheres trace back to Tsuji (1969) and
Auman (1969). The inclusion of sophisticated physics became available with the
work of Mould (1975, 1976) and has steadily continued with Allard (1990), Kui
(1991), Brett \& Plez (1993), Allard \& Hauschildt (1995), Brett (1995a,
1995b), Tsuji, Ohnaka \& Aoki(1996), Hauschildt et al. (1999).
 
We use the most recent grid of model atmosphere publicly available at the
Phoenix project's
website\footnote{ftp.hs.uni-hamburg.de/pub/outgoing/phoenix/GAIA/}. The models
cover a range of parameters far wider than that needed for the present work:
$2000 \le \Teff \le 10000\,\rm{K}$, $-0.5 \le \log (g) \le 5.5$ and $-4.0 \le
\textrm{[M/H]} \le +0.5$. Below $7000\,\rm{K}$ the grid is given in steps of 
$100\,\rm{K}$ in
effective temperature and 0.5~dex in metallicity. The molecular line lists
include about 700 million molecular lines, 15--300 million of which are
typically selected in a model. The equation of state is an extension of that
used in Allard \& Hauschildt (1995). For the coolest models the dust is assumed
to form and to immediately rain out completely below the photosphere (``cond'' 
models) so that it does not contribute to the opacity. Full details
are available in Brott \& Hauschildt (2005) and references therein.

Since we are working with dwarf stars, we assume $\log (g) = 5.0$
throughout. This assumption is in agreement with the values of $\log (g)$
determined from other techniques (S\'egransan et al. 2003; Berger al. 2006;
Bean et al. 2006a). As we will see later, a change of $\pm 0.5$~dex in the
assumed surface gravity implies only minor differences in the derived 
parameters.

The Phoenix models also include variations of $\alpha$--elements for each
metallicity. We have chosen to use models with no $\alpha$--enhancement; in any
case the use of $\alpha$--enhanced models in dwarfs of earlier spectral type
does not change the results significantly (Casagrande et al. 2006). There are
indications that M dwarfs follow the same [$\alpha$/Fe] vs. metallicity as
measured in FGK dwarfs (Woolf \& Wallerstein 2005) and since our sample is
limited to the solar neighborhood we do not expect significant signs of
$\alpha$--enhancement.

\subsection{Colour--Colour plots}\label{colcol}

Testing of synthetic model atmospheres is normally done by comparing observed
and modelled spectral energy distributions for a range of wavelengths and
spectral types (e.g. Tinney, Mould \& Reid 1993; Brett 1995b; Leggett et al. 
1996, 2000, 2001; Burgasser, Cruz \& Kirkpatrick 2007). Here we simply compare
synthetic and empirical colours, showing that with good accuracy data right
across the optical and infrared, photometry provides an excellent tool to test
model atmospheres (e.g. Bessell, Castelli \& Plez 1998). The interested reader 
can find e.g.\ in Leggett et al. (2000, 2001) and Burgasser et al. (2007) a 
thorough discussion of the comparison and pitfalls between observed and 
modelled spectra for M dwarfs and subdwarfs.
\begin{figure*}
\begin{center}
\includegraphics[scale=0.6]{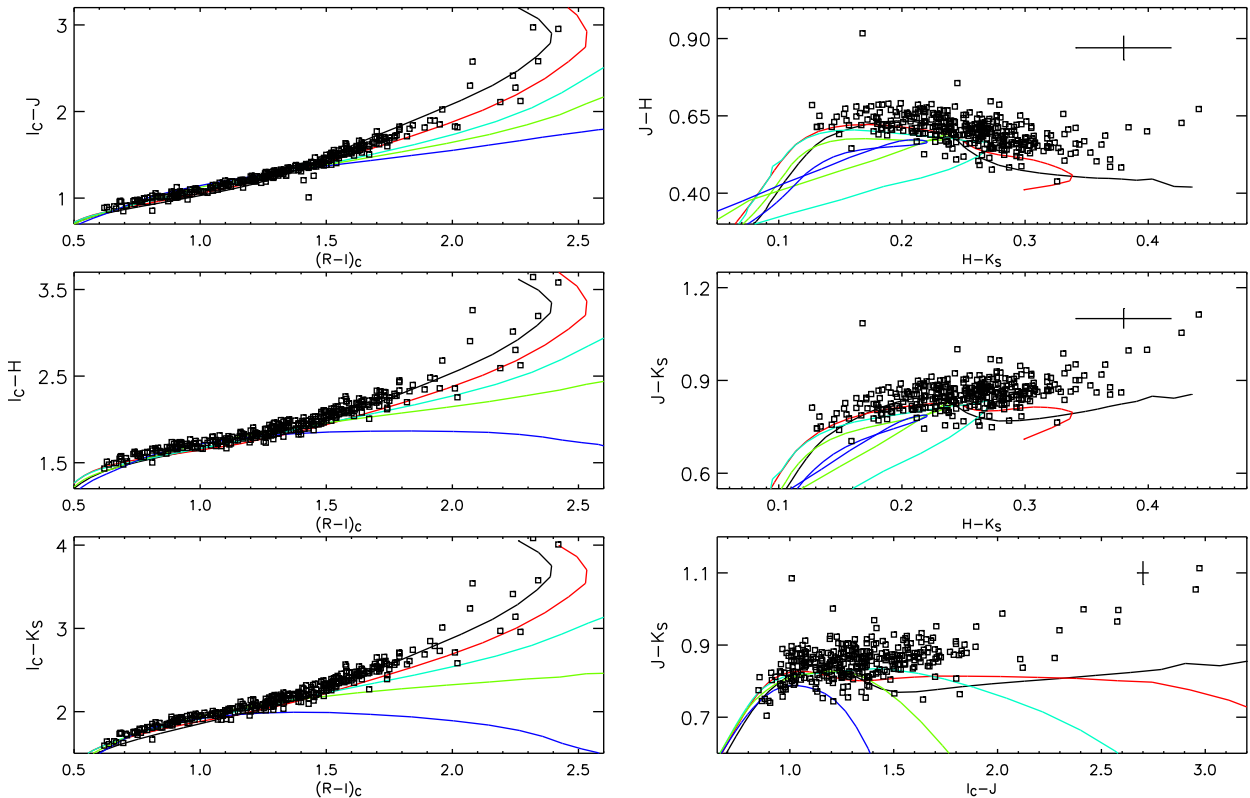}
\caption{Same as Figure \ref{vkplot} but as function of other colour indices.} 
\label{ijhkplot}
\end{center}
\end{figure*}

Synthetic colours are computed from the model atmospheres using prescriptions
very similar to those in Casagrande et al. (2006), as we discuss in more
details in Appendix B.
In Figure \ref{vkplot} we plot various colour indices as function of $V-K_S$.
It is clear from the figure that the models have problems in $U-B$, being
offset with respect to the observations by $0.1-0.2$~mag. Such an offset is
unlikely to be entirely due to uncertainties in standardize the $U$ band 
filter (see Appendix B) and/or the zero-points (as these amount to be at
most $0.01-0.02$~mag) and most likely reflects inadequacies in modeling the UV
spectral region in cool stars. The models behave considerably better in the
other optical and infrared bands, although $J-H$ and $J-K_S$ appear offset when
going to dwarfs redder then $V-K_S \gtrsim 5.5$ which corresponds to 
$\Teff \lesssim 3000\,\rm{K}$ (compare with Figure \ref{coltemp}).

Alternatively to $V-K_S$, another excellent temperature indicator in cool stars
is $I_C-J$.
In fact, Figure \ref{ijplot} looks very similar to Figure \ref{vkplot}, but 
the stars are now distributed over a shorter baseline. 

In the left panel of Figure \ref{ijhkplot} we show infrared indices as 
function of $(R-I)_C$,
which is sensitive to decreasing spectral type for M dwarfs (Bessell
1991). The models are in fair agreement with the data, but get worse going 
to the reddest $(R-I)_C$. Although at the very red end observed colours might 
be slightly less accurate (Section \ref{jc}), Bessell (1991) has shown that
the large spread in observed colours for the latest M dwarfs is real. 
The reason is likely to be that at such cool temperatures M dwarfs show the 
effects of dust at bluer wavelengths and have slightly different spectra 
with stronger hydride bands and weaker TiO and VO bands.
In the purely infrared colour planes of Figure \ref{ijhkplot} we are working 
in narrow colour ranges and
observational errors become prominent in the comparison. To help establish the
trends we have also checked how these planes look when our sample of M dwarfs 
is complemented with the earlier one for G and K dwarfs from 
Casagrande et al. (2006).  
At bluer colours the data show a
turnover in $J-H$ and $J-K_S$ as function of decreasing temperature
(i.e. increasing $H-K_S$ and $I_C - J$). The models partly predicted this
feature, which is the result of the sensitivity of infrared colours to the
photospheric gas pressure (Mould 1976) as well as to the occurrence of 
$\rm{H}_2\rm{O}$ bands. At very low effective temperatures, the
data suggest a flattening and a possibly a rising in $J-H$ and $J-K_S$, whereas
the models decrease steadily. The rising of the observed colours in the 
$H-K_S$ vs.\ $J-H$ plane is confirmed by similar plot using dwarfs much cooler 
than we have here (e.g. Reid et al. 2001; Burgasser et al. 2007).
This mismatch between data and models was
already noticed in other models by Brett (1995b), and essentially means that in
the models $H$ and $K_S$ magnitudes become progressively fainter than $J$
magnitude as the effective temperature decreases.

Overall, the Phoenix models show fair agreement with the data in various 
bandpasses, although inadequacies still persist, especially at the coolest 
temperatures where the dust needs to be properly incorporated. 
The synthetic colours also show a large spread with metallicity for 
decreasing $\Teff$: since our sample is limited to the solar neighbourhood, 
we expect our M dwarfs share a distribution similar to that observed in GK 
dwarfs, i.e. with most of the metallicities between $-0.5$ and $0.0$. 
To this respect, the coolest models are considerably offset with respect to 
the position of the stars which are encompassed by super-solar and solar 
lines, rather than solar and sub-solar as one would expect from the argument 
mentioned above.  
The serious discrepancies in $U$ band have little impact on the present 
work of calibrating the bolometric luminosities of the stars, since so 
little flux is emitted in the $U$ band. For the reddest stars,
which correspond to effective temperatures below $3000-2900\,\rm{K}$ (see Figure
\ref{coltemp}) the models do show some problems, especially in the infrared.
Since our technique partly relies on model atmospheres, this means that at the 
very cool end the results we present in Section \ref{abs_cor} to \ref{tesca} 
are still open to refinements. Looking at Figures \ref{vkplot} to 
\ref{ijhkplot} is obvious that by simply comparing observed and theoretical 
colours one would deduce different stellar parameters (effective temperatures 
and metallicities) depending on the colour index adopted for the comparison. 
The technique we present in the next Section is less 
affected by such inconsistencies. In fact, we will use model atmosphere 
only to estimate the flux outside our multi--band photometry, i.e. the flux in 
the blue and red tails of the spectra. This 
estimate should be rather accurate as long as model atmosphere reproduce 
the overall spectral energy distribution, even though specific bands might 
have problems. Notice though, where theoretical models do show 
limitations, like in the optical and near-infrared, we use anyway the observed 
colours. In addition, for estimating 
the metallicities we will calibrate our technique on other empirical 
measurements (see Section \ref{abs_cor}), and this should keep under control 
the deficiencies that still affect theoretical models. 
Nonetheless, it would be too optimistic to believe we are not affected by the 
inaccuracies in the models. Future improvements on the theoretical side will 
certainly benefit to our technique, especially below $3000\,\rm{K}$.

\section{The Multiple Optical--Infrared Technique}\label{mfm}

In our previous paper we used multi--band photometry to implement the IRFM and
we derived effective temperatures, bolometric luminosities and angular
diameters for a set of G and K dwarfs (Casagrande et al. 2006). It is
therefore natural to ask whether the same technique can be successfully applied
to M dwarfs.

Although the underlying idea of the IRFM is still valid when going to effective
temperatures cooler than $\sim 4000\,\rm{K}$, some caveats exist. Here, we 
generalize
and extend our temperature scale to dwarf stars much cooler than in Casagrande
et al. (2006) to which the reader can refer for an introduction to the
formalism and details on the computation of the bolometric and monochromatic
flux from multi--band photometry. The extension of the method presented here
concerns the computation of $\Teff$, which is now done by using the fluxes in
both optical and infrared bands (Section \ref{top_level}). 
For this reason we call our method the Multiple Optical--Infrared Technique 
(MOITE). As in Casagrande et al. (2006), the effective temperatures we derive 
depend on very few basic assumptions, namely the adopted Vega absolute 
calibration and zero-points (see also Appendix B). The dependence on the 
adopted grid of model atmospheres is also not so crucial since most of the 
bolometric flux (usually around 80 percent) is recovered from our 
multi--band photometry. The MOITE proves 
to be also sensitive to metal content in M dwarfs, as we discuss in Section 
\ref{met}.

\subsection{The IRFM in brief}

The basic idea of the IRFM (Blackwell \& Shallis 1977; Blackwell,
Shallis \& Selby 1979; Blackwell, Petford \& Shallis 1980) is to compare the
ratio between the bolometric flux $\mathcal{F}_{Bol} \textrm{(Earth)}$ and the
infrared monochromatic flux $\mathcal{F}_{\lambda} \textrm{(Earth)}$, both
measured at Earth (the so called observational $R_{obs}$ factor) to the ratio
between the surface bolometric flux ($\sigma \Teff^4$) and the surface infrared
monochromatic flux $\mathcal{F}_{\lambda}\textrm{(model)}$, predicted from
model atmospheres. The ratio of the last two quantities defines the theoretical
$R_{theo}$ factor. From this ratio $\Teff$ can be computed iteratively as
follow:

\begin{equation}\label{eq_irfm}
T_\mathrm{eff,n} = \left( \frac{\mathcal{F}_{\lambda}\textrm{(model)}_{(n-1)}
\mathcal{F}_{Bol}\textrm{(Earth)}_{(n-1)}}{\sigma
\mathcal{F}_{\lambda}\textrm{(Earth)}_{(n-1)}}\right)^{\frac{1}{4}}
\end{equation} 

where the effective temperature determined at the $n^{th}$ iteration depends on
the effective temperature determined at the $(n-1)^{th}$ iteration and which is
used to improve the estimate of the quantities on the right hand side of
equation (\ref{eq_irfm}). In the IRFM, more than one infrared band is usually
used (i.e. $R_{theo}$ is computed for each band), and the procedure
described here is applied to each band separately. At each iteration the
average $T_\mathrm{eff,n}$ obtained from all the infrared bands is then
computed and the procedure is iterated until the effective temperature 
converges to a final value.

The IRFM has been traditionally applied to stars of K or earlier spectral 
type, to derive effective temperatures approximately above $4000\,\rm{K}$.
Qualitatively, above this temperature, spectra roughly behave like black--body
curves in the infrared, so that in this region the spectra can be
described by the Rayleigh--Jeans law and the ratio between the bolometric and
monochromatic flux depends on the effective temperature to some power, with
little or no metallicity dependence (Figure \ref{fr}).

\subsection{From the IRFM to the MOITE: a top level description of the 
technique}\label{top_level}

We now generalize the technique presented in the previous Section to effective 
temperatures cooler than $4000\,\rm{K}$ using both optical and infrared bands 
with the MOITE.
We aim to give here a qualitative description of the underlying idea, leaving 
the full technical details to Appendix A.
\begin{figure*}
\begin{center}
\includegraphics[scale=0.55]{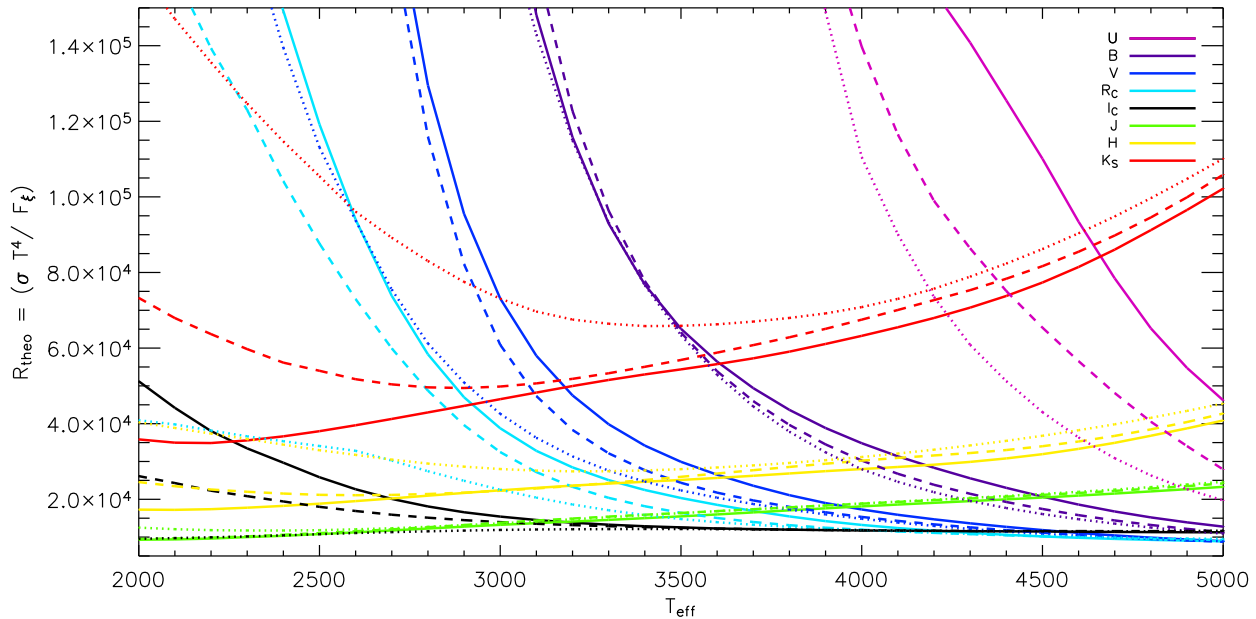}
\caption{Ratio between the bolometric and the monochromatic flux
($R_{theo}$) in various filter bandpasses $\xi$ (lines of different colours) as
function of $\Teff$ for the Phoenix models in three different metallicities 
$+0.0$~(solid), $-1.0$~(dashed) and $-2.0$~(dotted). 
Broad-band (heterochromatic) fluxes have been reduced to monochromatic fluxes 
as described in the Appendix B of Casagrande et al. (2006).}
\label{fr}
\end{center}
\end{figure*}

Below approximately $4000\,\rm{K}$, molecular absorption and flux 
redistribution become very important and significantly change the original
continuum shape, making any type of qualitative black--body description to an M
dwarf spectrum hazardous also in the infrared. Further, as the effective 
temperature decreases the
peak of the spectra moves redward, until at $\Teff \sim 3000\,\rm{K}$ it 
settles in $J$ band and stops moving further to the red (Allard \& Hauschildt 
1995). Below $\sim 4000\,\rm{K}$, depending on the infrared band, 
$R_{theo}$ flattens out and then monotonically increases with decreasing 
effective temperature. This behaviour in the infrared resemble that shown by 
$R_{theo}$ also in the optical bands (Figure \ref{fr}). 

It seems therefore obvious that below $4000\,\rm{K}$, depending on the 
metallicity, particular care must be used in determining the effective 
temperature by means of the flux ratio. On the other hand, since in cool 
stars $R_{theo}$ behaves qualitatively the same in both infrared and optical 
colours, once a technique for determining $\Teff$ is found, that can be 
readily applied to any photometric band. 

As we discuss in more detail in Appendix A, when $R_{theo}$ monotonically 
increases with decreasing temperature it is still possible to converge in 
$\Teff$ if we compare the observed flux product $\mathcal{F}_{Bol}
\textrm{(Earth)} \times \mathcal{F}_{\lambda} \textrm{(Earth)}$ to its
theoretical counterpart $\left( \frac{\theta}{2} \right)^{4} \sigma \Teff^4
\mathcal{F}_{\lambda}\textrm{(model)}$. The apparent drawback of this method is
that it introduces a dependence on the angular diameter, whereas such a
dependence was canceled out when doing the flux ratio. However at each $n^{th}$
iteration the angular diameter can be estimated from the $(n-1)^{th}$ iteration

\begin{equation}
\left( \frac{\theta}{2} \right)_{n}^{4} = \left( \frac{\mathcal{F}_{Bol}
\textrm{(Earth)}_{(n-1)}}{\sigma T_\mathrm{eff,n-1}^4} \right)^{2}
\end{equation}

so that it is still possible to converge in effective temperature

\begin{equation}\label{eq_mfm}
T_\mathrm{eff,n}=\left( \frac{\sigma \mathcal{F}_{\lambda} \textrm{(Earth)}_{(n-1)}
T_\mathrm{eff,(n-1)}^{8}} {\mathcal{F}_{Bol} \textrm{(Earth)}_{(n-1)}
\mathcal{F}_{\lambda}\textrm{(model)}_{(n-1)}} \right)^{\frac{1}{4}}.
\end{equation}

Since this formalism is valid when $R_{theo}$ monotonically increases with
decreasing temperature, the advantage of this approach is that now it is
possible to use also the optical colours to converge in effective temperature
below $\sim 4000\,\rm{K}$. Notice that to bootstrap either the IRFM or the 
MOITE one needs to interpolate over a grid of synthetic spectra according to 
the details given in Appendix A. To do so, the metallicity of a star must 
be known: when this is not possible, [M/H] has been obtained with the 
procedure we describe in the next Section.

\subsection{Estimating the metallicities of M dwarfs with the MOITE}\label{met}

Figure \ref{fr} shows that going to cooler $\Teff$, both optical and infrared
colours start to show a strong dependence on the metallicity. We exploit this
particularity to implement a novel technique to estimate the metallicities of
the M dwarfs.
 
The method works as follows: for a given star of unknown metallicity, 
we apply the MOITE to recover the effective temperature, assigning each time
a trial [M/H] to the star, from $-2.1$~dex to $0.4$~dex, in steps of
0.1~dex. The trial metallicity assigned to the star is used for interpolating 
over the grid of model atmospheres. 
The chosen metallicity range well brackets our (rather local) sample of M
dwarfs.  For a given star we obtain 6 $T_{\xi}$ which estimates $\Teff$ from
each of the colour bands individually $V(RI)_C JHK_S$, for each of the 
26 different metallicity choices (from $-2.1$ to $0.4$~dex). Since each band 
has a different
sensitivity to the metallicity, the scatter among the 6 $T_{\xi}$ is at a
minimum when the correct metallicity is chosen, as we prove in the next Section
for a set of synthetic colours.  
\begin{figure*}
\begin{center}
\includegraphics[scale=0.55]{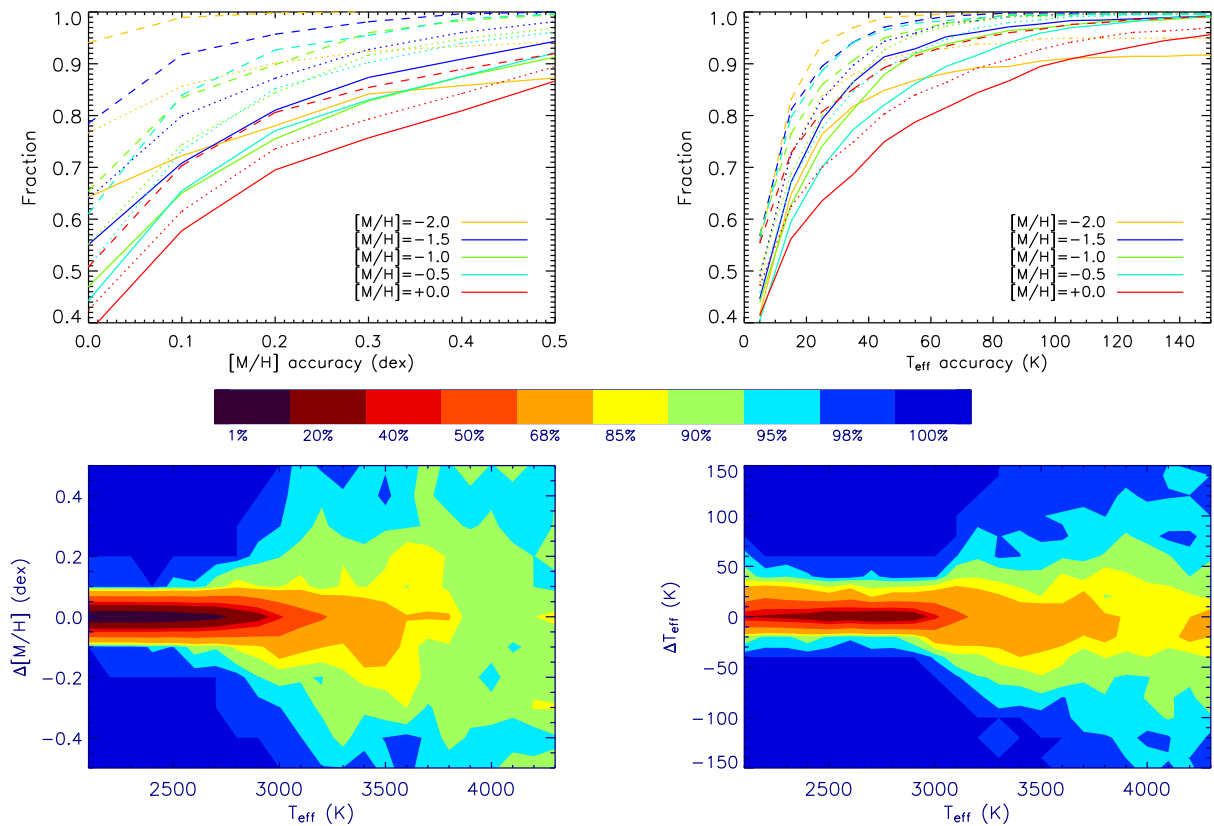}
\caption{Upper panels: accuracy of the MOITE in recovering the correct [M/H]
and $\Teff$ when realistic observational uncertainties (Section \ref{intacc}) 
are included in the
Monte Carlo simulation. Different line-styles correspond to the recovered
parameters for $\Teff \le 4300\,\rm{K}$ (continuous line), $3900\,\rm{K}$ 
(dotted) and $3500\,\rm{K}$ (dashed). Lower panels: contour plots showing 
the accuracy of the
method as function of $\Teff$. When going to cooler effective temperature the
fluxes in different bands become extremely sensitive to the metal content and
the physical parameters of the underlying synthetic spectra are always
recovered with excellent accuracy, as discussed in the text.}
\label{confidence}
\end{center}
\end{figure*}

For real data, there is an additional
complication. Empirically, for the 118 M dwarfs with known [M/H], we find that
the temperature estimates in each band $T_{\xi}$ are on average offset by a 
few 10s of K from the average $\Teff$ with a dependence on both [M/H] and 
$\Teff$. This offset might
be ascribed to zero-point errors in the absolute calibration (see also fig. 8
and 12 in Casagrande et al. 2006) and/or to systematics in the spectral
library, or both. The computation of $\mathcal{F}_{Bol}\textrm{(Earth)}$ and
$\mathcal{F}_{\lambda}\textrm{(Earth)}$ from the observed multi--band
photometry depend on the adopted zero-points and absolute calibration
(Casagrande et al. 2006). Although our adopted Vega absolute calibration has 
been thoroughly tested in both optical and infrared bands via ground 
(Tokunaga \& Vacca 2005) and spaced based (Bohlin \& Gilliland 2004; Price 
et al. 2004; Bohlin 2007) measurements, uncertainties at the level of a 
few percent are 
present and are almost certainly responsible for the systematic offsets of 
order tens of Kelvin in $T_{\xi}$ between different bands. Stars with different
metallicity and effective temperature emit differently in a
given $\xi$ band; since the adopted absolute calibration and zero-points
change the contribution of each $\xi$ band into the final result, this explains
why the offsets are function of [M/H] and $\Teff$.  We use the 118 M dwarfs
for which we know their metallicities to correct these offsets in $T_{\xi}$ 
obtained from each band. For the real stars, this reduces the scatter in 
temperatures for each of the trial metallicities, and considerably assists 
in the recovery of the correct metallicity.
Notice that this correction to properly estimate [M/H] for our stars is 
calibrated on the Bonfils et al. (2005) metallicity scale, but for a given 
metallicity, $\Teff$ is obtained with the MOITE alone. 

\begin{table*}
\centering
\caption{Observable and physical quantities for our sample stars.}
\label{TempBolMet}
\begin{tabular}{lcccccccccccc}
\hline
name &$\Teff \pm \Delta \Teff$&$\theta \pm \Delta \theta$ & $m_{Bol}$ & $V$ & $B-V$ & $U-B$ & $V-R_C$ & $V-I_C$ &    $J$  &   $H$   &  $K_S$ & [M/H] \\             
     &        (K)               &       (mas)               &           &     &       &       &         &         &         &         &        &       \\ 
\hline
HIP112         & $3923 \pm  142$ & $0.209 \pm 0.016$ &   9.668 &  10.748 &   1.410  &  1.320  &  0.879  &  1.722  &  8.017  &  7.408  &  7.217 & $-0.08$ \\ 
HIP897         & $3786 \pm  126$ & $0.228 \pm 0.016$ &   9.628 &  10.822 &   1.463  &  1.249  &  0.914  &  1.826  &  7.976  &  7.314  &  7.119 & $-0.18$ \\
HIP1734        & $3397 \pm  154$ & $0.306 \pm 0.029$ &   9.459 &  11.133 &   1.508  &  1.162  &  1.009  &  2.209  &  7.674  &  7.052  &  6.785 &  $0.07$ \\
HIP1842        & $3327 \pm  110$ & $0.241 \pm 0.017$ &  10.073 &  11.886 &   1.523  &  1.166  &  1.042  &  2.326  &  8.259  &  7.640  &  7.375 &  $0.04$ \\
\ldots    &  \ldots    &   \ldots  & \ldots  & \ldots   & \ldots  &   
\ldots    &  \ldots    &   \ldots  & \ldots  & \ldots   & \ldots  & \ldots \\ 
\hline
\end{tabular}
\begin{minipage}{1\textwidth}
Effective temperatures ($\Teff$) and angular diameters ($\theta$) are computed 
as described in the text. For the stars with HIP number, the
metallicities are from the Bonfils et al. (2005) calibration, while the
remainder are obtained with the MOITE. We only give metallicities for stars
with $\Teff$ above $3080\,\rm{K}$ for the reasons explained in Section 
\ref{abs_cor}.
Apparent bolometric magnitudes ($m_{Bol}$) are obtained according to Section
\ref{luca}, where the absolute bolometric magnitude of the Sun
$M_{Bol,\odot}=4.74$. Optical colours are in the Johnson-Cousins system, 
whereas infrared are from 2MASS. The full table is available in electronic 
format.
\end{minipage}
\end{table*}

\subsection{MOITE : accuracy of the technique}\label{intacc}

The first test for the MOITE is to ensure that the proposed technique works, 
and if so, to which accuracy. The best way to address the level of internal 
accuracy of the method is by using synthetic colours to check whether the 
correct physical parameters (i.e. $\Teff$ and [M/H]) of the underlying 
synthetic spectra are recovered.

We use the Phoenix model atmospheres to generate a set of synthetic $BV(RI)_C
JHK_S$ magnitudes for stars in the temperature range $2100 \le \Teff \le
4500\,\rm{K}$ and metallicity range $-2.0 \le \textrm{[M/H]} \le 0.0$~dex. 
Notice
that now the adopted absolute calibration and zero-points are not responsible
for any offset among $T_{\xi}$ in different bands, since the same absolute
calibration and zero-points are used in generating synthetic magnitudes and in
the MOITE.

We begin by applying the MOITE, regarding [M/H] as a fixed, known parameter.
To get the iteration started, initial temperature estimates were made using
Bessell's (1991) $\Teff:(R-I)_C$ calibration. We find that we can recover the
correct effective temperatures of the model spectra with an accuracy of 
$1 \pm 3\,\rm{K}$ and
the bolometric luminosities (i.e. $\sigma \Teff^4$) within 0.1 percent. We
then tested what happens if rather than using the Bessell (1991)
$\Teff:(R-I)_C$ calibration for the first estimate of the effective temperature
we start from either $\Teff=5000\,\rm{K}$ or from $2000\,\rm{K}$. 
The method still recovers
the correct temperatures and bolometric luminosities with the same accuracy as
before (but more iterations are needed). The MOITE is thus pretty insensitive
to poor first guesses of the effective temperature and always correctly
converges.

We have then tested the MOITE at recovering metallicity (i.e. introducing [M/H] 
as a free parameter) as well as effective
temperature and luminosity. We find we can recover the metallicities of the
underlying synthetic spectra with an accuracy of $0.006 \pm 0.04$~dex, with
very few cases when they deviate by 0.1~dex. As a consequence, $\Teff$ and the
bolometric luminosities are still recovered with very good accuracy. These
tests establish that the technique has a high internal accuracy in the ideal 
case of no observational errors.

For a more realistic approach, we mimic real data by running Monte Carlo
simulations with realistic observational uncertainties. For a set of synthetic
$BV(RI)_CJHK_S$ colours, we have assigned each time random errors with a normal
distribution centered on the synthetic values and a standard deviation equal to
the typical optical (Section \ref{jc}) and infrared (Section \ref{nir})
photometric errors.
The results are shown in Figure \ref{confidence}, and demonstrate that the
method becomes more accurate when going to lower $\Teff$, since at cool
temperatures the flux ratios show a pronounced dependence on the metallicity
as expected from Figure \ref{fr}. 
Now that realistic observational errors are taken into account we recover the 
metallicities of the synthetic spectra within $0.1-0.3$~dex and the effective 
temperatures within $50-100\,\rm{K}$, also depending on the spectral type. 
It is important to remember that real data might include systematic 
uncertainties (especially in the absolute calibration) which are difficult to 
assess and therefore not included when running the Monte 
Carlo simulations. The actual accuracy might be somewhat worse than in 
Figure \ref{confidence}.
Also, it is fair to remember that the comparison with the observed colours
(Section \ref{colcol}) shows that at coolest temperatures there is still room 
for improvements in theoretical modeling. For very cool stars, our results are
therefore subject to possible refinements.  However, even in the range 
$3000 < \Teff < 4000\,\rm{K}$ the method recovers [M/H] and $\Teff$ of the 
underlying synthetic spectra with an accuracy of $\sim 0.2$~dex and 
$\sim 50\,\rm{K}$ at a confidence level of $85$ percent (i.e. within 
$1.5 \; \sigma$).

We conclude that the MOITE shows a high degree of internal accuracy 
over the range of metallicities and effective temperatures covered in the 
present study. 
However, for a better estimate of the reliability of such a technique, we 
need to compare our results with direct measurements. This will be done in 
Section \ref{abs_cor} and \ref{tesca} where we will compare the metallicities 
and the effective temperatures returned by the MOITE with those recently 
measured in literature by means of other techniques.

\subsection{The final error budget}\label{erbu}

We have shown our technique to be very promising in obtaining both effective 
temperatures and metallicities of cool stars. 
Based on the comparison with other measurements, in the next Section we 
will estimate the uncertainty of our metallicities to be on average 
$0.2-0.3$~dex especially when photometry is available in many bands. 

For evaluating the final errors in effective temperatures, bolometric
luminosities and angular diameters, we use the same prescriptions as in
Casagrande et al. (2006). Briefly, for each star we run 200 Monte Carlo
simulations assigning each time random errors to the photometry (according to
the uncertainties given in Section \ref{jc} and \ref{nir}) and to [M/H].  We
have also accounted for a change of $\pm 0.5$~dex in the value of $\log (g)$
used in our grid of model atmospheres and for the case the errors in the
absolute calibration correlate to give systematically higher or lower
fluxes. When using the MOITE to estimate the metallicities, we search for the
solution that minimize the scatter among $T_{\xi}$. Although the scatter is
minimized, it is always finite: this reflects photometric errors as well
as possible systematics which are not fully corrected, especially at the lowest
temperatures. We adopt a very conservative approach and also include the
scatter in different bands in the final $\Teff$ uncertainty.

The effective temperatures, $m_{Bol}$ and the metallicities for our entire
sample are given in Table \ref{TempBolMet}.
On average, our effective temperatures are accurate to about $100\,\rm{K}$ which
corresponds to $2-3$ percent in the studied temperature range. The uncertainty
in bolometric luminosities is usually between 3 and 4 percent and that in
angular diameters between 4 and 6 percent. Also, the fact that the MOITE
recovers the metallicities consistently with other determinations, although is
not a proof of the correctness of $\Teff$ is reassuring and indirectly confirms
the accuracy of the temperatures (Figure \ref{confidence}).

\section{The M dwarf metallicity scale}\label{abs_cor}

In the previous Section we have introduced how the MOITE works and we have 
evaluated its accuracy by means of Monte Carlo simulations. 
Here, we compare the metallicities obtained with our technique to those 
recently measured by Bonfils et al. (2005), Woolf \& Wallerstein (2005, 2006) 
and Bean et al. (2006a, 2006b). For dwarfs with $\Teff$ below $3000\,\rm{K}$ 
we use {\it Hipparcos} common proper--motion companions 
to compare the metallicity we derive for the cool secondaries with that more 
easily measured for the hotter primaries in the aforementioned studies. 
We will compare our effective temperatures with other determinations available 
in the literature in Section \ref{tesca}.

\subsection{Accuracy of the metallicities above $3000\,\rm{K}$}\label{upT}

\begin{figure}
\begin{center}
\includegraphics[scale=0.55]{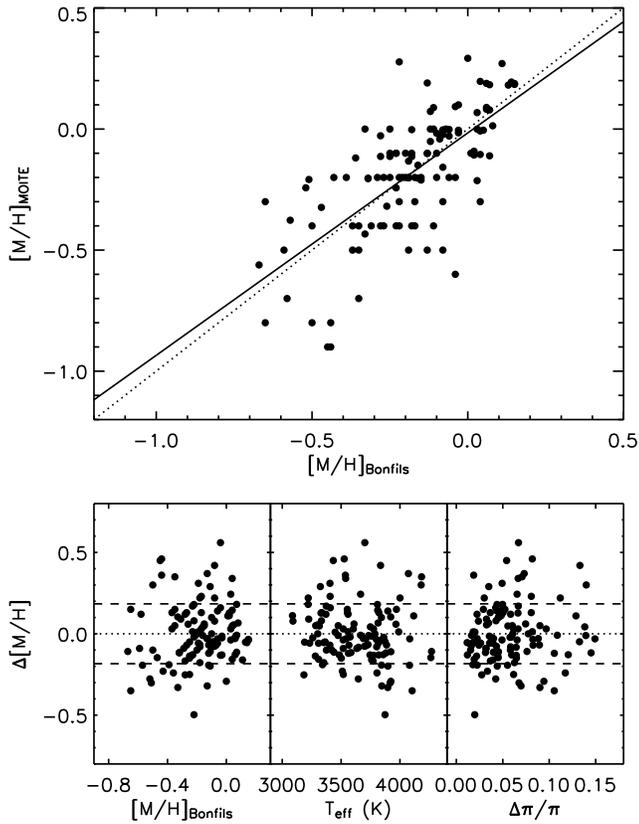}
\caption{Upper panels: comparison between the Bonfils et al. (2005)
metallicities and those obtained with the MOITE. The continuous line is the fit
to the data, the dotted line is the one--to--one relation. Lower panels: the
metallicity difference ($ \textrm{[M/H]}_{\textrm{\tiny{Bonfils}}} -
\textrm{[M/H]}_{\textrm{\tiny{MOITE}}} $) as function of other
parameters. Long--dashed lines are the $1 \sigma $ scatter.}
\label{moite}
\end{center}
\end{figure}

Figure \ref{moite} shows our metallicity estimates for the 118 M dwarfs with 
known metallicities, and shows a $1 \sigma$ scatter of 
0.2~dex, i.e. within the accuracy of the Bonfils et al. (2005) calibration.  
This comparison is encouraging, but still includes stars which were used to 
construct the calibration (Section \ref{met}), so is not a fully external 
check on the method.

\begin{figure*}
\begin{center}
\includegraphics[scale=0.55]{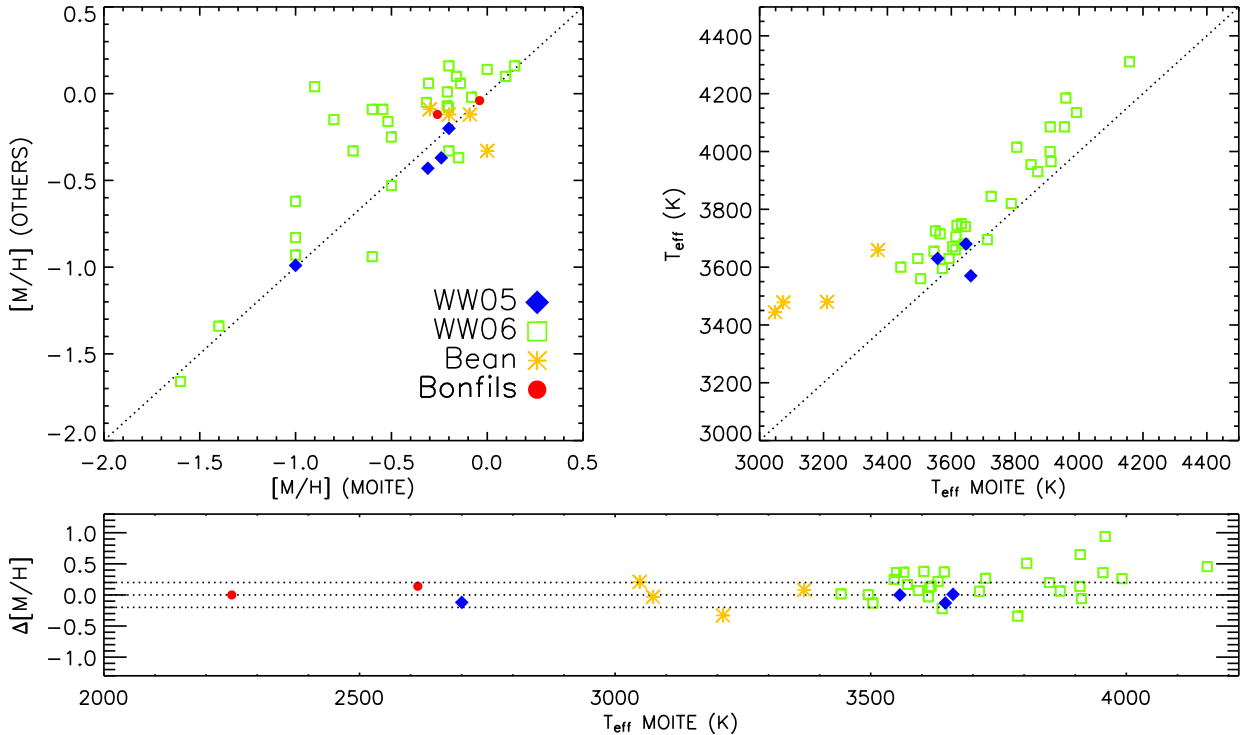}
\caption{Comparison between $\Teff$ and [M/H] obtained by the MOITE and those
measured from Woolf \& Wallerstein (2005) (filled diamonds), 
Woolf \& Wallerstein (2006) (open squares), Bean et al. (2006a, 2006b) 
(asterisks) and Bonfils et al. (2005) (filled circles) according to the 
details given in Section \ref{upT} and \ref{downT}. Notice that below 
$3000\,\rm{K}$ the metallicities are those measured in the {\it Hipparcos} 
common proper--motion companions. In the first and second panel, dotted 
diagonal lines with slope $1$ are intended to guide the eye. 
The horizontal dotted lines in the third panel highlight the $\pm 0.2$~dex 
interval around the mean zero. $\Delta \rm{[M/H]}$ refers to the
difference $\rm{[M/H]}_{\rm{others}} - \rm{[M/H]}_{\rm{MOITE}}$.}
\label{ww06}
\end{center}
\end{figure*}

Recently, reliable metallicities for M dwarfs have been measured by Woolf \&
Wallerstein (2005, 2006) from very high--resolution spectra. Spectral
synthesis technique has been also successfully applied by Bean et al. (2006a,
2006b). We have extensively searched for $BV(RI)_C JHK_S$ photometry of M
dwarfs analyzed in the aforementioned studies and found accurate measurements
for those reported in Table \ref{ww}: for $\Teff$ approximately above 
$3000\,\rm{K}$ the mean difference in metallicity is just $-0.03 \pm 0.06$~dex 
($\sigma=0.17$~dex). In particular, 
the comparison with the direct spectroscopic measurements of Woolf \& 
Wallerstein (2005, 2006) agrees always within $0.13$~dex (filled 
diamonds in Figure \ref{ww06}). The agreement with the [M/H] measurements of 
Bean et al. (2006a, 2006b) is somewhat poorer (asterisks in Figure \ref{ww06}), 
but there are large discrepancies between their and our $\Teff$ scale, as we 
discuss later in this Section. 
Such large differences in the adopted $\Teff$ obviously reflect in the
abundances measured by Bean et al (2006a, 2006b).

\begin{table*}
\centering
\caption{Magnitudes and colours for various M dwarfs and the $\Teff$
and [M/H] recovered with the MOITE compared with those obtained by other
techniques. In the infrared we have used $JHK_S$ magnitudes from 2MASS.}
\label{ww}
\begin{tabular}{lcrrrrcccccc}
\hline
Name & NLTT & $V$  & $B-V$  & $V-R_C$  & $V-I_C$ & Ref. & $\Teff^{\textrm{\tiny{MOITE}}}$~(K)  & $\rm{[M/H]}_{\textrm{\tiny{MOITE}}}$ & $\Teff$~(K) & $\rm{[M/H]}_{\rm meas.}$ & Ref. \\
\hline
GJ 191     & 14668 &  8.841 & 1.570 & 0.956 & 1.951 & 1 & 3661 &$ -1.00$ & 3570 &$ -0.99$ & a\\
GJ 701     & 45883 &  9.362 & 1.515 & 0.976 & 2.060 & 2 & 3557 &$ -0.20$ & 3630 &$ -0.20$ & a\\
GJ 828.2   & 51282 & 11.090 & 1.534 & 0.964 & 1.967 & 2 & 3646 &$ -0.24$ & 3680 &$ -0.37$ & b\\
GJ 876     & 55130 & 10.179 & 1.571 & 1.182 & 2.733 & 2 & 3076 &$ -0.09$ & 3478 &$ -0.12$ & c\\
GJ 581     & 39886 & 10.568 & 1.602 & 1.109 & 2.501 & 2 & 3211 &$ \phantom{-} 0.00$ & 3480 &$ -0.33$ & c\\
GJ 297.2 B & 19072 & 11.80  & 1.49  & 1.03  & 2.29  & 4 & 3370 &$ -0.20$ & 3659 &$ -0.12$ & d\\
GJ 105   B & 8455  & 11.66  & 1.50  & 1.22  & 2.78  & 4 & 3048 &$ -0.30$ & 3444 &$ -0.09$ & d\\
GJ 412 B   & 26247 & 14.44  & 2.08  & 1.66  & 3.77  & 3 & 2700 &$ -0.31$ & $-$  &$ -0.43$ & a\\
GJ 618   B & 42494 & 14.15  & 1.79  & 1.412 & 3.233 & 3 & 2614 &$ -0.26$ & $-$  &$ -0.12$ & e\\
GJ 752   B & 47621 & 17.20  & $-$   & 2.10  & 4.36  & 3 & 2250 &$ -0.04$ & $-$  &$ -0.04$ & e\\
\hline
\end{tabular}
\begin{minipage}{1\textwidth}
Source of the optical photometry : (1) Kilkenny et al. (1998); (2) Koen et
al. (2002); (3) Bessell (1990a); (4) Laing (1989).  Source of metallicity : (a)
Woolf \& Wallerstein (2005); (b) Woolf \& Wallerstein (2006); (c) Bean et
al. (2006b); (d) Bean et al. (2006a); (e) Bonfils et al. (2005). For the
Bonfils et al. (2005) calibration we have used the $V$ magnitudes given in the
Table and the $K_S$ magnitudes from 2MASS.
\end{minipage}
\end{table*}

Notice that our technique relies on the metallicities determined from Bonfils et
al. (2005), whose calibration also includes measurements from Woolf \&
Wallerstein (2005). To further and independently test our results, we have
applied the MOITE to all the stars in Woolf \& Wallerstein (2006). For these
stars, only $VJHK_S$ photometry is available. We have used $V$ magnitudes as
given in table 1 of Woolf \& Wallerstein (2006) and $JHK_S$ from 2MASS. Again,
we have used only stars with total photometric errors in the infrared smaller
than 0.10 mag. Since we are now running the MOITE using fewer colours, we might
expect our results to be slightly less accurate. The comparison with the 
Woolf \& Wallerstein (2006) measurements in Figure \ref{ww06} is reassuring, 
with a mean difference of $0.20 \pm 0.05$~dex ($\sigma = 0.27$~dex) and it 
validates our technique over a larger range of metallicities. 

Differently from the study of FGK dwarfs, M dwarfs exhibit complex spectra 
which render far less trivial the determination of both effective temperatures 
and metallicities from spectroscopic analysis only. 
The purpose of this Section is to evaluate the accuracy of our metallicities, 
whereas we compare our temperature scale with other existing ones 
in Section \ref{tesca}. 
It is however worth discussing here the differences with the effective 
temperatures adopted by Woolf \& Wallerstein (2005, 2006) and Bean et al. 
(2006a, 2006b) to derive their metallicities.  
Our effective temperatures are systematically cooler than those of Woolf \& 
Wallerstein (2006), especially for earlier M spectral types ($3700\,\rm{K}$ 
and above), where the difference in metallicities is also higher. 
Woolf \& Wallerstein (2005, 2006) estimate $\Teff$ from theoretical 
color-temperature relations in $V-H$ and $V-K_S$ obtained using older 
model atmospheres (Hauschildt et al. 1999). According to Woolf \& 
Wallerstein (2006), systematics as high as $100-200\,\rm{K}$ in their 
$\Teff$ could not be excluded. Our effective temperatures are in 
agreement with the latest Phoenix and above $3500\,\rm{K}$ Castelli \& 
Kurucz (2003) models, as it can be seen from Figure \ref{coltemp}. 
The disagreement in $B-V$ for the Castelli \& Kurucz (2003) models is likely 
due to pitfall in accounting the contribution of all molecular features in 
these spectral regions.
The effective temperatures
obtained by the spectral fitting technique of Bean et al. (2006a, 2006b) 
rely particularly on the TiO bandhead. Bean et al. (2006a) also notice that 
while at M0.5 spectral types their $\Teff$ agree with the scale of Reid \& 
Hawley (2005), their effective temperatures increase linearly with spectral 
type and for their latest M dwarf (GJ 105 B) they are about $300\,\rm{K}$ 
hotter than Reid \& Hawley (2005) and $400\,\rm{K}$ hotter then our scale. 
This hotter temperature scale is possibly related to drawback in determining 
effective temperatures via spectral synthesis of certain bandhead 
(e.g. Jones et al. 2005).
We will further discuss and test our temperature scale with other recent 
determinations, in particular with interferometric angular diameters, in 
Section \ref{tesca} and prove our effective temperatures to be reliable.

Concluding, the metallicities estimate with the MOITE above $\sim 3000\,\rm{K}$ 
agree within $0.2-0.3$~dex to those measured by other recent studies, 
especially when all $BV(RI)_CJHK_S$ photometry is available.
In particular, the comparison with direct spectroscopic measurements of 
Woolf \& Wallerstein (2005) suggest an even better agreement, at the level of 
$0.1$~dex.

\subsection{Accuracy of the metallicities below $3000\,\rm{K}$}\label{downT}

As we have already mentioned, the calibration performed in Section \ref{met} 
is function of $\Teff$ and [M/H] and in our case is obtained in the 
metallicity and temperature range shown in Figure \ref{moite}. The 
comparison in Figure \ref{ww06} suggests that the technique can be safely 
applied to metallicities as low as about $-2.0$~dex, at least above 
$3500\,\rm{K}$ and to metallicities typical of the solar neighbourhood down 
to about $3000\,\rm{K}$.
Here, we would like to test to what extent we can use the MOITE to measure 
metallicities in stars with effective temperatures below this latter limit.
The extrapolation to lower temperatures is particularly
interesting, since on the basis of our Monte Carlo simulations the MOITE is 
expected to recover the metallicity to high accuracy (always within 
$0.1-0.2$~dex) below $3000\,\rm{K}$ (Figure \ref{confidence}). Also, for 
the coolest effective temperatures, as we discussed in Section \ref{colcol}, 
theoretical models do show discrepancies with respect to the empirical 
colour--colour diagrams. Those could reduce the formal high accuracy 
expected for $\Teff < 3000\,\rm{K}$, so that below this temperature it is 
interesting to compare our with respect to other more ``direct'' 
metallicity estimates.  

We are not aware of any [M/H] measurement for M dwarfs with 
$\Teff < 3000\,\rm{K}$, but from the list of Gould \&
Chanam\'e (2004), we have searched for multi--band optical and infrared
photometry of late M dwarfs which are common proper--motion companions to
\emph{Hipparcos} stars. We have then compared our metallicities estimated 
with the MOITE to those more easily measured for the primary \emph{Hipparcos} 
stars. We have found three late M
dwarfs (GJ 412 B, GJ 618 B, GJ 752 B) whose common proper--motion companions
are early type M dwarfs with metallicities from Woolf \& Wallerstein (2005) or
from the calibration of Bonfils et al. (2005). We caution that some of the
stars in Table \ref{ww} are classified as ``variable'' in SIMBAD and GJ 618 B
has very large photometric errors in 2MASS (much larger than our usual fiducial
level ``j\_'', ``h\_'' and ``k\_msigcom'' $< 0.10$~mag). 
\begin{figure*}
\begin{center}
\includegraphics[scale=0.55]{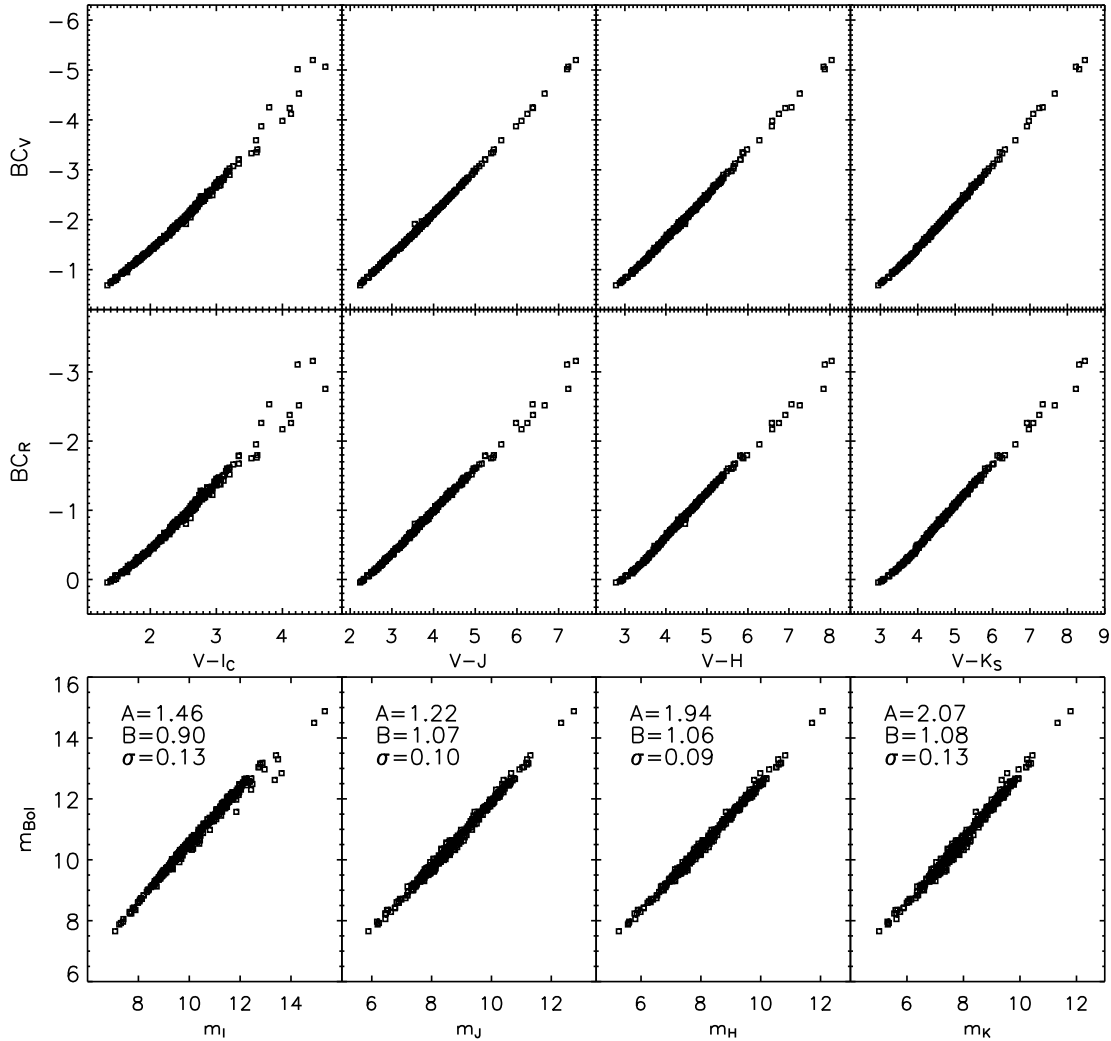}
\caption{Upper and central panel: bolometric corrections in $V$ and $R$ bands 
as function of various color indices for our M dwarfs. Lower panel: relation 
between the magnitudes observed in different bands $m_{\xi}$ and the bolometric 
magnitude $m_{Bol}$. Linear fits in the form 
$m_{Bol}=\textrm{A}+\textrm{B}\,m_{\xi}$ are given along with the resulting 
standard deviation $\sigma$.}
\label{bc}
\end{center}
\end{figure*}

There are only three stars and it is hard to draw definitive 
conclusions, but it is reassuring to see such a nice agreement, with the 
metallicity of the secondaries in agreement with that of the primaries 
within 0.14~dex even in the worst case. For two stars 
[M/H] of the primary is obtained using the Bonfils et al. (2005) formula, 
which we have used to calibrate our technique. 
Nonetheless, now we are working with effective
temperatures much cooler than those of Figure \ref{moite} and the metallicities
are still properly recovered. Of course more stars are needed before 
confidently extend our technique to such cool stars, but
the method looks promising. To be on the safe side, for our entire sample of 
M dwarfs
(Table \ref{TempBolMet}) we give [M/H] only for stars with $\Teff$ above
$3080\,\rm{K}$, i.e. only where the MOITE is safely calibrated as explained in
Section \ref{met}.

\section{Empirical bolometric corrections}\label{luca}

We adopt the same definition of Casagrande et al. (2006) to define the 
bolometric correction in a given $\xi$ band, where
\begin{equation}\label{bolcomp}
\textrm{BC}_{\xi}=m_{Bol}-m_{\xi}
\end{equation}
and the zero-point of the $m_{Bol}$ scale is fixed by choosing 
$M_{Bol,\odot}=4.74$.
Although it is possible to give analytic transformations between the flux and
various colour indices, bolometric corrections can be readily computed using 
equation (\ref{bolcomp}) in any band from the stars in Table \ref{TempBolMet}, 
and are probably more useful. 

The bolometric correction in $V$ and $R$ bands as function of various colour 
indices are shown in Figure \ref{bc}. In the optical there is some dependence
on the metallicity among the coolest stars, but when going to longer colour
baselines, the data show very tight relations, especially in $V-J$. 
The flux emitted by cool stars peaks in the near--infrared. The 
bolometric corrections in $I_C$, $J$, $H$ and $K_S$ are almost constant as 
function of different color indices. For these bands it is therefore possible 
to pass directly from the observed $m_\xi$ to the bolometric magnitude $m_{Bol}$ 
via linear fit as shown in Figure \ref{bc} and given here: 
\begin{equation}
m_{Bol} = \left\{ \begin{array}{l}
       1.46 + 0.90 \,m_{I_{C}}\\
       1.22 + 1.07 \,m_{J}\\
       1.94 + 1.06 \,m_{H}\\
       2.07 + 1.08 \,m_{K_{S}}.
\end{array} \right.
\end{equation}
Of course, by using the data in Table 
\ref{TempBolMet} to fit the bolometric correction as function of a given colour 
index it is possible to achieve higher accuracy still. Nonetheless, the 
linear fits in Figure \ref{bc} are useful to have a quick estimate of the 
bolometric magnitude given the apparent one.
From these bolometric corrections and the colour--temperature relations given 
in the next Section it is also possible to obtain very accurate estimates of 
the angular diameter of M dwarfs, as we describe in Section \ref{inter}.

\section{The M dwarfs temperature scale}\label{tesca}

\begin{figure*}
\begin{center}
\includegraphics[scale=0.55]{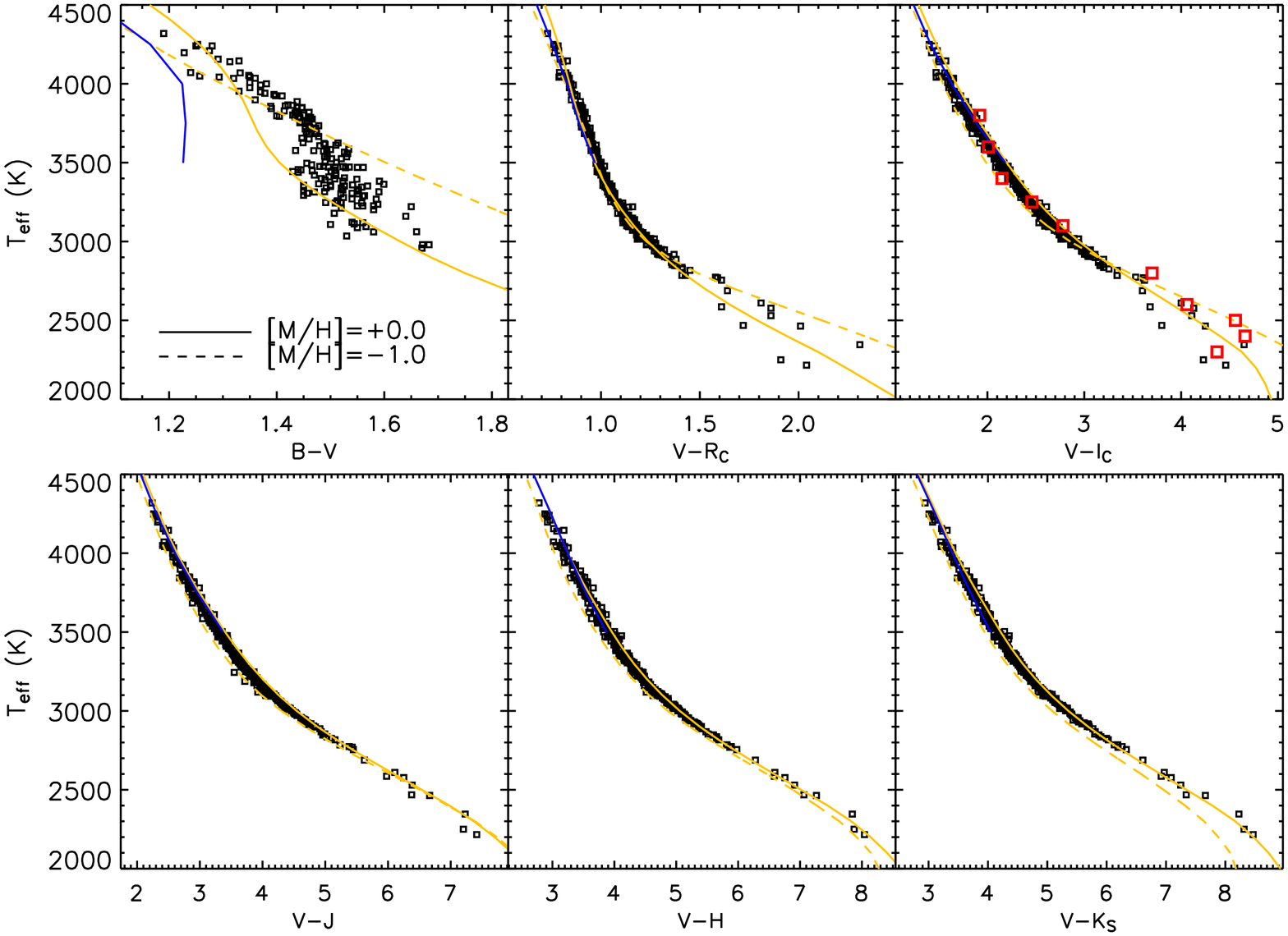}
\caption{$\textrm{Colour}-\Teff$ plots in different bands for our M dwarfs. 
Overplotted 
are the prediction from the Phoenix models (yellow lines) for two different 
metallicities which roughly bracket our sample of stars. Also shown for 
comparison the prediction from the Castelli \& Kurucz (2003) models for solar 
metallicity (blue line). Red squares in the $\Teff$ vs. $V-I_C$ plot are from 
the temperature scale of Reid \& Hawley (2005).}
\label{coltemp}
\end{center}
\end{figure*}

We have fitted the observed $\textrm{colour}-\Teff$ relations by analytical 
fits, using the following functional form:
\begin{equation}\label{thetarel}
\theta_{\mathrm{eff}}=a_0 + a_1 X + a_2 X^{2} + a_3 X^{3}
\end{equation}
where $\theta_{\mathrm{eff}}=5040 / \Teff$, $X$ represents the colour and 
$a_{i}\;(i=0,1,2,3)$ are the coefficients of the fit.
Depending on the band (Figure \ref{coltemp}), the data show an increasing 
scatter in the effective temperature estimates below $\sim 2800\,\rm{K}$, 
which is very likely due to metallicity dependencies. 
Because of this, we strongly caution against using $V-R_C$, $V-I_C$ and 
$(R-I)_C$ indices below $2800\,\rm{K}$. We have also verified that such 
scatter at low temperatures is not present in the other transformations of 
Table \ref{ctr}.
We do not have enough stars below this temperature to address the question 
further and the metallicity estimates are still somewhat uncertain so that 
we do not include any metallicity dependence in our analytic fits. 
Furthermore our M dwarfs are drawn from the solar neighbourhood, and they 
have a limited range of metallicities. It will be interesting to try the 
method on halo M dwarfs, as these become available especially in large 
photometric/spectroscopic surveys currently underway (e.g. from RAVE, SEGUE, 
SkyMapper) and later with GAIA.
We thus differ from the Casagrande et al. (2006) fitting formulae, which 
accounted for the larger metallicity coverage of that sample. Also, now 
we need to fit a third order polynomial to account for the inflection that 
occurs at lower $\Teff$. For this reason, the fitting formulae of Table 
\ref{ctr} are not an exact continuation of those in Casagrande et al. (2006), 
which are correct for $\Teff$ hotter than $4400\,\rm{K}$. 
Therefore, the colour ranges for the fits in Table \ref{ctr} do not
overlap with Casagrande et al. (2006), but are given for slightly redder 
colours. If a link between the two scales is needed, however, we advise the
users to a careful case by case study.

With the exception of $B-V$, none of the colour-temperature transformations 
have strong dependence on [M/H] above $2800\,\rm{K}$, and therefore our 
relations are not likely to be affected much as metallicities for M dwarfs 
improve.
The temperature fit to the $B-V$ colour has huge scatter, and gives only the 
crudest temperature estimate. If one really needs to estimate $\Teff$ from 
$B-V$ the best choice is probably to use Figure \ref{coltemp}. 

We have searched for DENIS photometry so as to give in Table \ref{ctr} 
the $\textrm{colour}-\Teff$ relations also in this system, although for a
smaller number of stars. According to the DENIS database, those magnitudes have
larger photometric errors (on average between 0.05 to 0.09 magnitudes) than 
2MASS so this might explain why the 
relations in Table \ref{ctr} are less accurate for the DENIS colours. Also, 
we have found relations involving the $I_{\textrm{\tiny{DENIS}}}$ very noisy and 
we do not give them. 

For most of the stars in Koen et al. (2002) SAAO $JHK$ photometry is also 
available from Kilkenny et al. (2007) and in Table \ref{ctr} 
colour--temperature relations are given in this system, too. 
Stars with SAAO infrared photometry are all hotter than $3000\,\rm{K}$ and 
because of the reduced temperature range, second order polynomial fits are 
accurate enough now.
We caution against extrapolating these relations to cooler temperatures, in 
particular for indices $I_C-J_{\textrm{\tiny{SAAO}}}$, 
$I_C-H_{\textrm{\tiny{SAAO}}}$ and $I_C-K_{\textrm{\tiny{SAAO}}}$.

\begin{table*}
\centering
\caption{Coefficients and range of applicability of our colour-temperature 
relations (Eq.\ \ref{thetarel}). 
The photometric systems are $V(RI)_C$ for Johnson--Cousins, 
$JHK_S$ for 2MASS, $(JK)_{\textrm{\tiny{DENIS}}}$ for DENIS and 
$(JHK)_{\textrm{\tiny{SAAO}}}$ for SAAO. For some indices, we caution the users 
from extrapolating these relations to redder colours than those given, as 
explained in Section \ref{tesca}.}
\label{ctr}
\begin{tabular}{lcrrrrrcccccc}
\hline
Colour & Colour range & $a_0$ & $a_1$ & $a_2$ & $a_3$ & $N$ & $\sigma(\Teff)$\\
\hline  
$V-R_C$  & [0.800, 2.310]  & $-1.4095$ & 5.1212     & $-2.7937$ & 0.5432    & 325 & 33  \\
$V-I_C$  & [1.400, 4.650]  & 0.5050    & 0.5562     & $-0.0593$ & 0.0027    & 333 & 26 \\
$V-J$    & [2.260, 7.231]  & 0.1926    & 0.5738     & $-0.0726$ & 0.0042    & 318 & 17 \\
$V-H$    & [2.946, 8.041]  & $-0.4711$ & 0.8450     & $-0.1161$ & 0.0066    & 315 & 23 \\
$V-K_S$  & [3.219, 8.468]  & $-0.4809$ & 0.8009     & $-0.1039$ & 0.0056    & 313 & 19 \\
$V-J_{\textrm{\tiny{DENIS}}}$ &[2.211, 7.124]& $-0.0386$& 0.7709     & $-0.1243$ & 0.0085    & 187 & 38 \\
$V-K_{\textrm{\tiny{DENIS}}}$ &[2.951, 8.306]& $-0.2756$& 0.6907     & $-0.0846$ & 0.0045    & 192 & 42 \\
$V-J_{\textrm{\tiny{SAAO}}}$ & [2.195,4.115]& 0.4445 & 0.3837 & $-0.0232$ & $-$ & 80 & 29 \\ 
$V-H_{\textrm{\tiny{SAAO}}}$ & [2.882,4.802]& 0.0406 & 0.4752 & $-0.0298$ & $-$ & 80 & 36 \\
$V-K_{\textrm{\tiny{SAAO}}}$ & [2.994,5.063]& 0.1609 & 0.3978 & $-0.0210$ & $-$ & 80 & 32 \\
$(R-I)_C$& [0.660, 2.270]  & 0.8326    & 0.6122     & $-0.0849$ & 0.0164    & 331 & 27 \\
$R_C-J$  & [1.503, 5.374]  & 0.3594    & 0.7223     & $-0.1401$ & 0.0134    & 329 & 19 \\
$R_C-H$  & [2.053, 6.001]  & $-0.1645$ & 0.9269     & $-0.1674$ & 0.0135    & 332 & 31 \\
$R_C-K_S$& [2.212, 6.428]  & $-0.0570$ & 0.7737     & $-0.1226$ & 0.0091    & 326 & 25 \\
$R_C-J_{\textrm{\tiny{DENIS}}}$&[1.481, 5.214]& 0.1541   & 0.9537    & $-0.2183$ & 0.0215 & 165 & 41 \\
$R_C-K_{\textrm{\tiny{DENIS}}}$&[2.221, 6.396]& $-0.2094$& 0.8900    & $-0.1495$ & 0.0109 & 160 & 43 \\
$R_C-J_{\textrm{\tiny{SAAO}}}$& [1.434, 2.949]& 0.6269 & 0.4385 & $-0.0334$ & $-$ & 81 & 32 \\
$R_C-H_{\textrm{\tiny{SAAO}}}$& [2.121, 3.638]& 0.1666 & 0.5733 & $-0.0466$ & $-$ & 81 & 41 \\
$R_C-K_{\textrm{\tiny{SAAO}}}$& [2.233, 3.899]& 0.2913 & 0.4615 & $-0.0299$ & $-$ & 81 & 35 \\
$I_C-J$  & [0.865, 2.954]  & $-0.3813$ & 2.6488     & $-1.1642$ & 0.1981    & 329 & 37 \\
$I_C-H$  & [1.433, 3.644]  & $-2.5844$ & 4.3925     & $-1.5386$ & 0.1941    & 338 & 77 \\
$I_C-K_S$& [1.592, 4.085]  & $-1.8798$ & 3.0706     & $-0.9024$ & 0.0989    & 339 & 61 \\
$I_C-J_{\textrm{\tiny{SAAO}}}$& [0.796, 1.428] & 0.1346    & 1.6307 & $-0.4078$   & $-$ & 81 & 45\\
$I_C-H_{\textrm{\tiny{SAAO}}}$& [1.483, 2.118] & $-1.6423$ & 2.6765 & $-0.5318$ & $-$ & 81 & 85 \\
$I_C-K_{\textrm{\tiny{SAAO}}}$& [1.595, 2.379] & $-0.8748$ & 1.6962 & $-0.2678$ & $-$ & 81 & 61 \\
\hline
\end{tabular}
\begin{minipage}{1\textwidth}
Notes. $N$ is the number of stars employed for the fit after the $3 \sigma$
clipping and $\sigma(\Teff)$ is the final standard deviation (in Kelvin) of the
proposed calibrations.
\end{minipage}
\end{table*}

Many M dwarfs are intrinsically variable, owing to spots or other activity. We
have cleaned the sample from variable stars as best we can, (and indeed our
colour--temperature relations are very tight), however there might still be
unrecognized long term variables present. The relations we give thus 
apply also to intrinsically low variability stars. 

For the sake of completeness, we have applied the MOITE to the M dwarfs labeled
as variable in the original Koen et al. (2002) sample and which were excluded,
according to the selection criteria of Section \ref{sample}. The stars were
found to follow the same mean locii of the colour--temperature relations as the
non-variables, but with a larger scatter, usually about twice the 
$\sigma(\Teff)$ of Table \ref{ctr}. One should keep in mind
that the stars in Koen et al. (2002) were already preselected in {\it
  Hipparcos} in order to exclude extremely active stars: how well our results
would apply to these latter objects we leave to future studies.

The latest Phoenix model atmospheres show good agreement with our
colour--temperature relations in Figure \ref{coltemp}. 
This partly reflects the fact that our scale has
been obtained using these models in the MOITE, but we are using a great deal of
observational information to recover the total bolometric luminosity, and the
model atmospheres are used only to estimate the missing flux, which is at most
of order twenty percent. We demonstrate further the reliability of our 
temperature scale in what follows: in Section \ref{inter} using 
recent determinations of angular diameters for M dwarfs and in Section 
\ref{theothers} comparing our results with those obtained by 
various recent temperature studies.

\subsection{MOITE versus interferometric angular diameter 
  measurements}\label{inter}

Although in the past, much work has been done in determining the effective
temperature scale of the M dwarfs, a firmly established scale has not been
achieved. Until recently, in fact, the only two M dwarfs with measured linear
diameters were the eclipsing binaries YY Gem (Kron 1952, Habets \& Heintze
1981) and CM Dra (Lacy 1977, Metcalfe et al. 1996), but the limiting factor in
accurately determining their effective temperatures were the parallaxes.
Long--baseline interferometry has recently provided angular diameters
measurements for an handful of nearby M dwarfs which can be used to test the
accuracy of our effective temperature and bolometric luminosity scale
(S\'egransan et al. 2003; Berger et al. 2006). Unlike G and K dwarfs, for which
all the interferometric targets have saturated 2MASS photometry, half of the M
dwarfs with measured angular diameters have good 2MASS colours.

The stellar angular diameters obtained with the MOITE are computed from the 
basic definition
\begin{equation}\label{teta}
\mathcal{F}_{Bol}\textrm{(Earth)} = \left( \frac{\theta}{2} \right)^{2} 
\sigma \Teff^{4}
\end{equation}
so that in principle a conspiracy of wrong effective temperatures and 
bolometric luminosities could still return correct angular diameters. 
However, the bolometric luminosities of our 
targets are observed via multi--band photometry (only subject to minor 
corrections to estimate the missing flux, see also Appendix A), so that 
$\mathcal{F}_{Bol}\textrm{(Earth)}$ is fixed by the observations and therefore 
comparison of our angular diameters with those measured by interferometers 
automatically tests our temperature scale.

We caution that even interferometric angular diameter measurements depend
mildly on modelling assumptions, in particular the limb-darkening corrections
to convert the measured uniform--disk angular diameters into the physical
limb-darkened disks ($\theta_{\rm LD}$) and to which we compare our 
$\theta$ of equation (\ref{teta}). The
limb-darkening coefficients used for M dwarfs (Claret 2000) are computed using
solar abundance atmospheric models, whereas the interferometric targets of
Table \ref{ang} span a larger metallicity range.  Another source of
uncertainty is due to the fact that limb-darkening coefficients are calculated
using 1D atmospheric models, whereas 3D models predict a less significant
center--to--limb variation. Such a difference might be up to a few percent in 
$\theta_{\rm LD}$ for hotter F and G stars, but is expected to be much smaller 
in the case of M dwarfs (Allende Prieto et al. 2002; Bigot et al. 2006). Since
all interferometric measurements reported here have been performed in the
infrared, where limb-darkening effects are minimized, we expect these 
uncertainties to be within the observational errors.

\begin{table*}
\centering
\caption{Comparison between the MOITE effective temperatures and angular
diameters (columns 8 and 9) and the interferometric measured ones (columns 10
and 11).}
\label{ang}
\begin{tabular}{lcrrrrclclcc}
\hline
Name           & $V$   & $U-B$ & $B-V$ & $V-R_C$& $V-I_C$&Ref.& $\Teff$~(K)   & $\theta_{\rm MOITE}$& $\Teff$~(K)   & $\theta_{\rm LD}$   & Ref. \\
\hline
GJ 191  & 8.841 & 1.186 & 1.570 &  0.956 & 1.951  & 1 & $3661 \pm 77$  & $0.637 \pm 0.028$ & $3570 \pm 156$ & $0.692 \pm 0.060$ & s\\
GJ 205 $\dag$         & 7.968 & 1.183 & 1.475 &  0.972 & 2.055  & 2 & $3546 \pm 106 $& $1.093 \pm 0.066$ & $3520 \pm 170$ & $1.149 \pm 0.110$ & s\\
GJ 411 $\dag$         & 7.47  & 1.14  & 1.51  &  1.01  & 2.15   & 3 & $3467 \pm 104 $& $1.515 \pm 0.091$ & $3570 \pm 42$  & $1.436 \pm 0.030$ & s\\
GJ 514  & 9.05  & $-$   & 1.52  &  0.98  & 2.04   & 4 & $3594 \pm 101$ & $0.636 \pm 0.037$ & $3243 \pm 160$ & $0.753 \pm 0.052$ & b\\
GJ 526 $\dag$         & 8.464 & $-$   & $-$   &  0.971 & 2.070  & 5 & $3533 \pm 106$ & $0.884 \pm 0.053$ & $3636 \pm 163$ & $0.845 \pm 0.057$ & b\\
GJ 699  & 9.553 & 1.264 & 1.737 &  1.228 & 2.779  & 6 & $3145 \pm 69$  & $1.003 \pm 0.046$ & $3163 \pm 65$  & $1.004 \pm 0.040$ & s\\
GJ 752 A& 9.115 & 1.138 & 1.515 &  1.039 & 2.333  & 2 & $3343 \pm 107$ & $0.835 \pm 0.054$ & $3368 \pm 137$ & $0.836 \pm 0.051$ & b\\
GJ 880  & 8.65  & $-$   & 1.497 &  0.985 & 2.103  & 4 & $3544 \pm 153$ & $0.822 \pm 0.072$ & $3277 \pm 93$  & $0.934 \pm 0.059$ & b\\
GJ 887 $\dag$         & 7.335 & $-$   & 1.500 &  0.975 & 2.02   & 4 & $3577 \pm 107$ & $1.411 \pm 0.085$ & $3626 \pm 56$  & $1.388 \pm 0.040$ & s\\
\hline
\end{tabular}
\begin{minipage}{1\textwidth}
Source of the optical photometry: (1) Kilkenny et al. (1998); (2) Koen et al. 
(2002); (3) Celis (1986); (4) Bessell (1990a); (5) The, Steenman \& Alcaino 
(1984); (6) Landolt (1983). In the infrared we have used 2MASS $JHK_S$
photometry (not reported here). A $\dag$ indicate a poor 
2MASS photometry, so that the angular diameter has been obtained from the
calibration of Section \ref{tesca} and \ref{luca} and not running the MOITE 
directly. Source of interferometric measurements: (s) 
S\'egransan et al. (2003); (b) Berger et al. (2006). 
\end{minipage}
\end{table*}

When running the MOITE for M dwarfs in Table \ref{ang} with good 2MASS 
photometry we have used the metallicities from Woolf \& Wallerstein
(2005) or applied the Bonfils et al. (2005) calibration, except for GJ 699 which
is outside of the Bonfils' et al. (2005) range of applicability and for which a
solar metallicity is thought to be appropriate (Leggett et al. 2000; Dawson \&
De Robertis 2004). The errors have been computed as described in Section
\ref{erbu}. For the other stars (i.e. those with inaccurate 2MASS photometry)
we apply the bolometric luminosity and effective temperature calibrations of 
Section \ref{luca} and \ref{tesca}. We estimate
$\Teff$ using the $V-I_C$ index, which has little intrinsic scatter above 
$2800\,\rm{K}$ (see Table \ref{ctr} and Figure \ref{coltemp}).
We then
compute the bolometric correction in $V$ band by linearly fitting the
$BC_V\;\rm{vs.} \; V-I_C$ relation (Figure \ref{bc}) 
in the colour range [1.95, 2.25].  As for
the temperature calibration, this relation has little intrinsic scatter and the
linear fit in the given range is accurate to 0.015~mag. Once the bolometric
correction and the effective temperature are known, the angular diameter can be
readily computed (eq. 14 in Casagrande et al. 2006). To these stars, we assign
a 3 percent error in $\Teff$ and 6 percent error in angular diameter,
consistently with the errors obtained for the other stars in the
sample. Although for these stars we are not applying the MOITE directly, we are
using the calibrations obtained with the MOITE itself so that they are fully
representative of our temperature and luminosity scale.

\begin{figure}
\begin{center}
\includegraphics[scale=0.55]{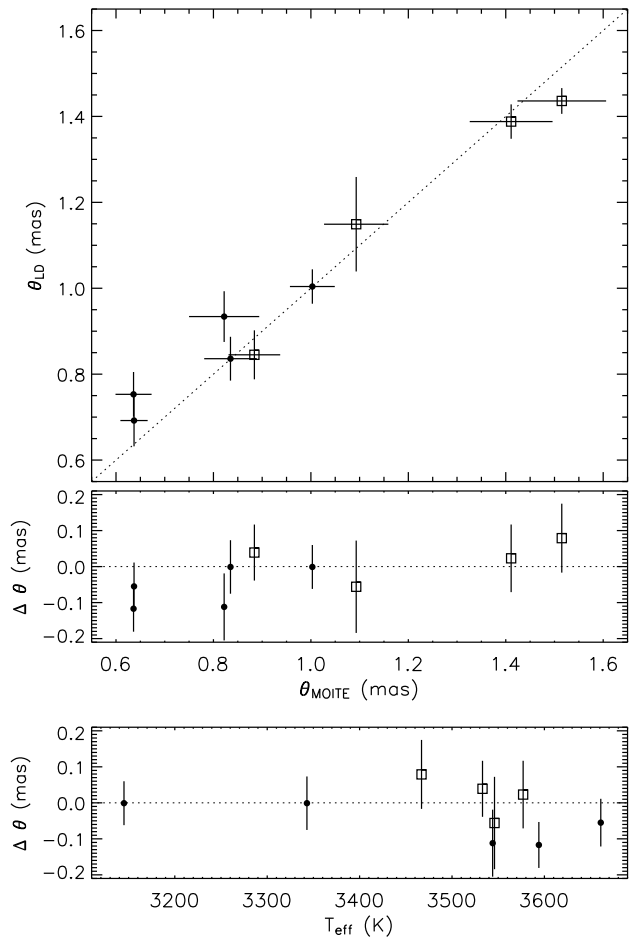}
\caption{Comparison between our and the interferometric $\theta_{\rm LD}$
angular diameters. Circles refer to angular diameters obtained applying the
MOITE directly. Squares are for those stars for which the angular diameters
have been computed from the colour--temperature and colour--luminosity
calibrations of Section \ref{tesca} and \ref{luca}. Dotted lines are intended
to guide the eye.}
\label{ivm}
\end{center}
\end{figure}

\begin{figure}
\begin{center}
\includegraphics[scale=0.5]{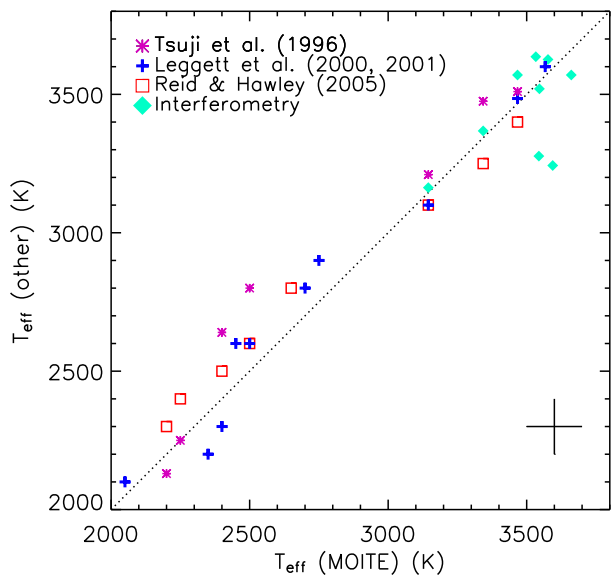}
\caption{Comparison between the MOITE effective temperatures and those 
determined in other recent studies. A typical error bar of $\pm 100\,\rm{K}$ 
is shown in the lower right corner. Dotted line with slope one is intended 
to guide the eye.}
\label{tempe}
\end{center}
\end{figure}

The overall comparison between our and the interferometric angular diameters 
shown in Figure \ref{ivm} is very good.  
S\'egransan et al. (2003) and Berger et al. (2006) also compute effective 
temperatures which are obtained combining the measured $\theta_{\rm LD}$ with 
bolometric correction polynomial fit or comparing observed and model-predicted 
fluxes based on the observed angular diameters. This allows a direct comparison 
with their effective temperatures in Figure \ref{tempe} (filled diamonds).
Only two stars deviate by more than $1 \sigma$ from the one--to--one relation 
in Figure \ref{ivm} and \ref{tempe}, namely GJ 514 and GJ 880. 
For these two stars, however, the two methods of determining $\Teff$ from 
$\theta_{\rm{LD}}$ (i.e. using bolometric 
correction or comparing observed and model-predicted fluxes) return results 
discordant by 
$100\,\rm{K}$. Choosing $\Teff$ computed using bolometric correction 
(table 4 in Berger et al. 2006) would reduce by $100\,\rm{K}$ the discrepancy 
for these two 
stars in Figure \ref{tempe}. We do not know the reason for such
disagreement, however, our bolometric corrections are carefully determined 
from multi--band data, whereas the main point of interferometric studies is to 
precisely measure angular diameters (with which we are in good agreement), 
rather than determining accurate empirical bolometric corrections. 
The important point from the comparison in this 
Section is that overall our angular diameters and effective temperatures 
agree with the trend defined by interferometric studies.

In particular, Barnard's star (GJ 699) is one of the benchmarks in setting the
cool star temperature scale. Our angular diameter is in excellent agreement
with the interferometric one and our effective temperature $\Teff = 3145 \pm
69$ also closely matches the value of $3134 \pm 102$, obtained by the careful
analysis of Dawson \& De Robertis (2004).

\subsection{Comparison with other temperature scales}\label{theothers}

The determination of effective temperatures by means of different techniques
below $4000\,\rm{K}$ becomes increasingly difficult as a result of the 
increasing complexity of the stellar spectra and previous studies have 
usually computed $\Teff$ for a limited number of stars.
Interferometry has recently provided a breakthrough to anchor the temperature 
scale down to $3000\,\rm{K}$, as discussed in Section \ref{inter}. Moreover, 
stars cooler than $3000\,\rm{K}$ have angular diameters too small to be 
resolved by currently available interferometers. Here we compare our 
effective temperatures to those obtained by various recent studies.

Among the possible different techniques to estimate $\Teff$, one is to fit
observed molecular features with model predictions (e.g. Kirkpatrick et al. 
1993). However, these features might depend not only on the effective 
temperature, but on other effects of line formation and the 
reliability of the models themselves. It is a long-standing result that 
especially below $3500-3000\,\rm{K}$ such a technique returns effective 
temperatures that are higher by several hundred kelvin with respect to other 
more empirically motivated methods (e.g. Pavlenko \& Jones 2002; Jones 
et al. 2005). Very recently Viti et al. (2008) have proposed a new and 
promising technique based on very high resolution mid-infrared observations 
of pure rotational water vapour transitions in M dwarfs.
Another approach is to compare observed
and synthetic spectra and estimate $\Teff$ from the model that better matches 
the observation in the infrared (e.g. Leggett et al. 2000, 2001), in the 
optical (e.g. Dawson \& De Robertis 2000) or throughout most of the spectrum 
(e.g. Pavlenko et al. 2006).  Burgasser \& Kirkpatrick (2006) have 
shown that the parameters derived using optical or near-infrared 
fits for a given object exhibit clear systematic differences up to 
$100-200\,\rm{K}$ in effective temperature and $0.5-1.0$~dex in metallicity. 

A more
consistent way to determine $\Teff$ is to analyze the entire
spectral region contributing to the bolometric flux, although in the past this
approach has been done mostly with black-body calibrations rather than with
model predictions (e.g. Veeder 1974; Reid \& Gilmore 1984).  A more rigorous
attempt to recover the bolometric flux was used by Tsuji et al. (1996). 
Finally, Reid \& Hawley (2005) have collected spectroscopic and photometric 
$\Teff$ estimates of a few well studied nearby M dwarfs covering the spectral 
type M0 to M9 (their table 4.1).

We have searched in the literature for other $\Teff$ determination of our 
stars and did not find many, particularly below $3000\,\rm{K}$. To increase 
the number of stars available for the comparison, we have applied our 
technique to a few more very red dwarfs (LHS36, LHS68, LHS292, LHS429, 
LHS2065, LHS2924, LHS3003) which are commonly studied in the literature. 
For these stars optical colours are available from Bessell (1990a, 1991) and 
infrared from 2MASS. We caution that they are all classified as flare stars 
in SIMBAD and for this reason they were not included in the original sample of 
Section \ref{sample}.
The effective temperatures for these and few other cool stars from Table 
\ref{TempBolMet}, \ref{ww} and \ref{ang} are discussed below. 

{\tt LHS2} - (GJ 1002). For this star we obtain $\Teff=2750\,\rm{K}$, slightly 
cooler than the temperature of $2900\,\rm{K}$ obtained by Leggett et al. (2000).

{\tt LHS36} - (GJ 406). We obtain an effective temperature of $2500\,\rm{K}$ 
which is cooler by about $100\,\rm{K}$ than the value of $2600\,\rm{K}$ 
obtained by Leggett et al. (2000) and reported also in Reid \& Hawley (2005). 
For the same star Pavlenko et al. (2006) obtained an effective 
temperature of $2800\,\rm{K}$ after a critical examination of the most recent 
model atmosphere fit to this object. The same temperature was also obtained 
by Tsuji et al. (1996). Golimowski et al. (2004) also found a hotter 
temperature than we do ($2900\,\rm{K}$). 

{\tt LHS37} - (GJ 411). For this star we obtain $\Teff=3467\,\rm{K}$ which is 
in agreement with $3510\,\rm{K}$ reported in Tsuji et al. (1996), 
$3500\,\rm{K}$ in Leggett et al. (1996) and $3400\,\rm{K}$ in Reid \& Hawley 
(2005).

{\tt LHS39} - (GJ 412 B). We obtain an effective temperature of $2700\,\rm{K}$,
which again is cooler by $100\,\rm{K}$ with respect to the value obtained by 
Leggett et al. (2000) ($2800\,\rm{K}$).

{\tt LHS57} - (GJ699). Our value $\Teff=3145\,\rm{K}$ is halfway between 
$3100\,\rm{K}$ in Leggett et al. (2000), Reid \& Hawley (2005) and 
$3210\,\rm{K}$ in Tsuji et al. (1996).

{\tt LHS65} - (GJ 821). For this {\it Hipparcos} star we obtain 
$\Teff=3567\,\rm{K}$, in close agreement with $3600\,\rm{K}$ in Leggett et 
al. (2000).

{\tt LHS68} - (GJ 866). This dwarf, together with GJ 406 is one of the 
reddest standards of the $BV(RI)_C$ system. Unfortunately it is member of a 
triple system, which might decrease the accuracy of the photometry 
(Delfosse et al. 1999). We obtain $\Teff=2650\,\rm{K}$, considerably 
cooler than the value of $3000\,\rm{K}$ obtained by Dawson \& De Robertis 
(2000) but in better agreement with $\Teff=2800\,\rm{K}$ in Reid \& Hawley 
(2005).

{\tt LHS292} - (GJ 3622). Our technique returns $\Teff=2450\,\rm{K}$, cooler 
then both Leggett et al. (2000) ($2600\,\rm{K}$) and Golimowski et al. (2004) 
($2725\,\rm{K}$).

{\tt LHS429} - (GJ 644 C). The MOITE returns $\Teff=2400\,\rm{K}$ for this 
late M dwarf, which is now $100\,\rm{K}$ hotter than in Leggett et al. (2000) 
but cooler than the value of $2500\,\rm{K}$ in Reid \& Hawley (2005) and 
$2640\,\rm{K}$ in Tsuji et al. (1996).

{\tt LHS473} - (GJ 752 A). The effective temperature we obtain $3343\,\rm{K}$ 
is midway between $3250\,\rm{K}$ in Reid \& Hawley (2005) and $3475\,\rm{K}$ 
in Tsuji et al. (1996).

{\tt LHS474} - (GJ 752 B). We obtain $\Teff = 2250\,\rm{K}$ which compares 
nicely with Tsuji et al. (1996) ($2250\,\rm{K}$) but it is slightly cooler 
than $2400\,\rm{K}$ in Reid \& Hawley (2005).

{\tt LHS2065} - (GJ 3517). For this very red dwarf we obtain 
$\Teff=2050\,\rm{K}$ in rough 
agreement with Leggett et al. (2001) ($2100\,\rm{K}$) but much cooler than 
the temperature of $2400\,\rm{K}$ obtained by Golimowski et al. (2004). 

{\tt LHS2924} - (GJ 3849). According to Reid \& Hawley (2005) this is the best 
studied M9 dwarf for which they report $\Teff=2300\,\rm{K}$. We obtain an 
effective temperature of $2200\,\rm{K}$ which is slightly hotter than the 
value of $2130\,\rm{K}$ in Tsuji et al. (1996). 

{\tt LHS3003} - (GJ 3877). We obtain $\Teff=2350\,\rm{K}$ which is now hotter 
than Leggett et al. (2001) ($2200\,\rm{K}$), but still cooler than 
Golimowski et al. (2004) ($2600\,\rm{K}$).

Our temperatures are shown in Figure \ref{tempe} against other studies 
discussed here or in Section \ref{inter}.
There is an overall good agreement with respect to the effective temperatures 
obtained by Leggett et al. (2000, 2001), especially above $3000\,\rm{K}$. 
Below this temperature typical differences of order $100\,\rm{K}$ exist, but 
on average the scatter in the data suggest we are on the same scale. 
Similarly, we are in very good agreement with the temperatures reported in 
Reid \& Hawley (2005) above $3000\,\rm{K}$, whereas below this temperature 
we are systematically cooler by $100\,\rm{K}$. We also agree within 
$100\,\rm{K}$ with Tsuji et al. (1996) except for effective temperatures 
around $2500\,\rm{K}$ where we have two stars with larger differences.
Our effective temperatures are also $250-400\,\rm{K}$ cooler than those in 
Golimowski et al. (2004) which are estimated using the relationship 
between the bolometric luminosity (their observable) and $\Teff$ predicted 
from evolutionary models.

Summarizing, further work is needed before reaching a consensus among different
temperature scales, especially below $3000\,\rm{K}$. However, our 
temperatures are supported from interferometric angular diameters between 
$3100$ and $3600\,\rm{K}$ and the homogeneous and smooth colour--temperature 
relations of Figure \ref{coltemp} lead us to believe our temperature scale is 
credible below $3000\,\rm{K}$. Also, the data in Figure \ref{tempe} 
suggest that despite 
case-by-case differences, on average we agree with the effective temperature 
reported in many recent studies.

\section{On the discontinuous transition from K to M dwarfs}\label{jump}

\begin{figure*}
\begin{center}
\includegraphics[scale=0.55]{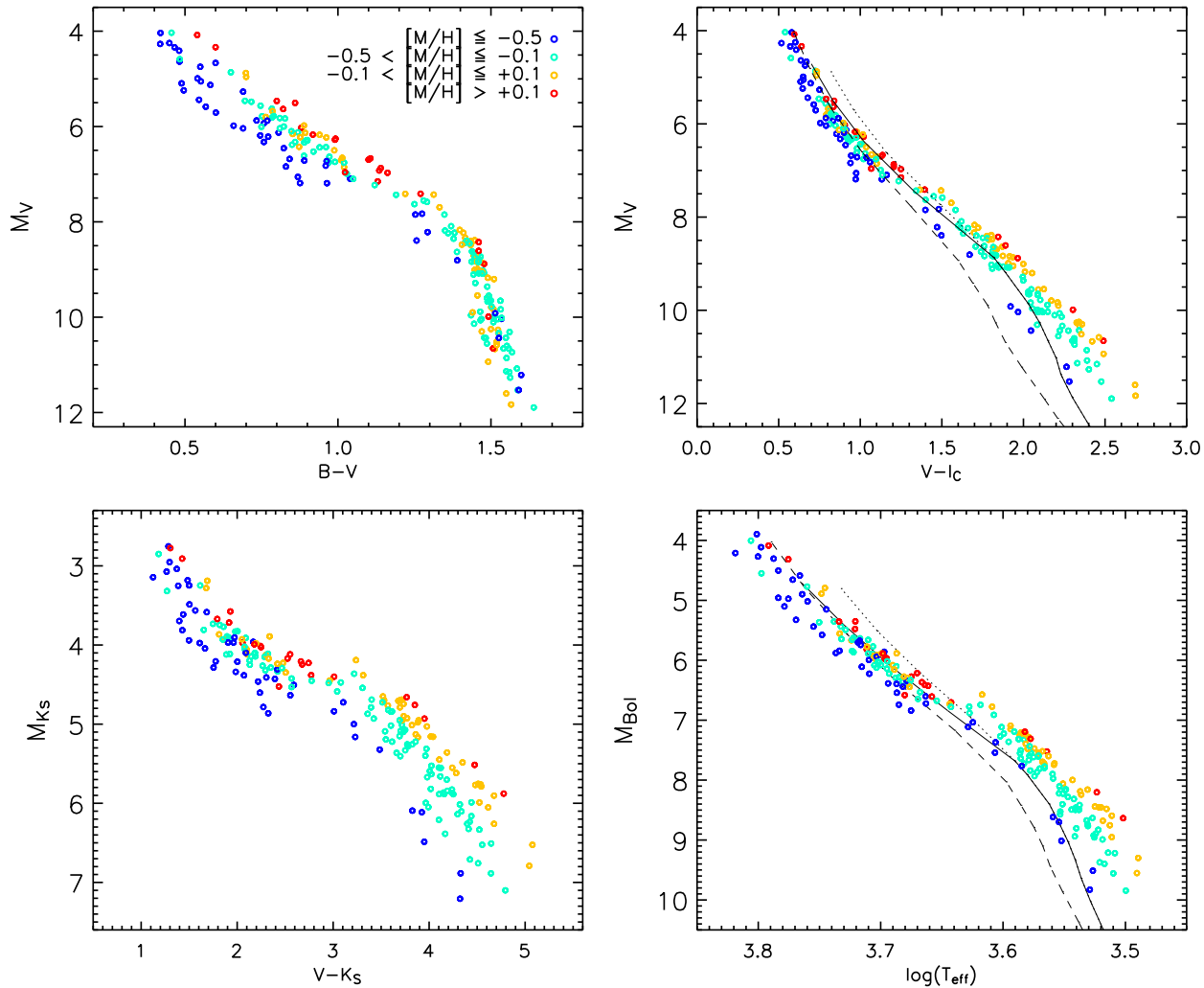}
\caption{HR diagram in different planes. Only stars with \emph{Hipparcos}
parallaxes better than 15 percent are used.  Overplotted are also the
theoretical isochrones from Baraffe et al. (1998) for $\rm{[M/H]}=0.0$ (dotted)
and $\rm{[M/H]}=-0.5$ (dashed) for $\alpha_{\rm MLT}=1$. The continuous line is
a solar calibrated model with $\alpha_{\rm MLT}= 1.9$. For all the isochrones
the age used is 5 Gyr. In the last panel both our sample of stars and the
theoretical isochrones have been plotted adopting $M_{Bol,\odot}=4.74$. Notice
that the metallicities for the stars plotted here are either from
high--resolutions spectroscopy (Casagrande et al. 2006) or from the Bonfils et
al. (2005) calibration.}
\label{bump}
\end{center}
\end{figure*}

In our previous paper we have implemented the IRFM to derive effective
temperatures and bolometric luminosities of G and K dwarfs. Here we have
extended our technique to much cooler effective temperatures. For M dwarfs with
accurate \emph{Hipparcos} parallaxes (i.e. those from Koen et al. 2002) we can
compute absolute magnitudes and therefore plot them with our previous sample
from Casagrande et al. (2006) and study the properties of the entire lower main
sequence (Figure \ref{bump}), finding a very interesting feature: whereas the
transition from late K to early M type occurs smoothly in the widely used 
$M_V-(B-V)$ plane, in infrared colours (which are better tracers of $M_{Bol}$ 
and
$\Teff$) a prominent discontinuity appears around late K to early M types. The
discontinuity is clearly evident in the observational plane $M_{K_S}-(V-K)$
(and we have verified it is present also in the other infrared colours) and is
also quite prominent in the theoretical $M_{Bol}-\Teff$ plane, at around
$4200-4300\,\rm{K}$ (Figure \ref{bump}).

The feature is close to the point at which our two calibrations (IRFM for G and
K dwarfs and MOITE for M dwarfs) meet. To confirm the discontinuity is not
dependent on the calibrations, we have reprocessed all the stars in Casagrande
et al. (2006) with the MOITE, and using the Phoenix model atmosphere,
confirming that we obtain the same temperatures and luminosities with both
methods above $\sim 4000\,\rm{K}$. 

For both the GK dwarfs in Casagrande et al. (2006) and the M dwarfs studied
here we have used strict selection criteria to remove double and variable
stars. For the sake of completeness, we note that when the M dwarfs labeled as
variable in Koen et al. (2002) are plotted in both the observational and
$M_{Bol}-\Teff$ planes of the HR diagram, these follow the same trend defined
from the {\it Hipparcos} stars of Section \ref{sample}.

We note that the discontinuity is as clear in the purely observational plane
most sensitive to temperature and luminosity ($M_{K_S}-(V-K)$) as it is in
the $M_{Bol}-\Teff$ plane: we thus consider the temperature-luminosity
discontinuity to be real. It occurs at $4200-4300\,\rm{K}$, appearing as a 
plateau in
the temperature-luminosity plane, with the luminosity of the M dwarfs holding
fairly steady even as their temperatures decrease. For this to occur, the radii
of the M dwarfs must be increasing again, rather than falling monotonically
going down the main sequence.

In Figure \ref{bump}, we compare our stars with the very low mass star
evolutionary models of Baraffe et al. (1998). We have adopted 5 Gyr old
isochrones, although the evolution of the lower main sequence is practically
insensitive to the age for $M_{Bol}>5.4$ (e.g. Casagrande et al.  2007). Figure
\ref{bump} shows that these models do not appear to reproduce the discontinuity
in the main sequence at $4200-4300\,\rm{K}$.

For further insight in the problem, in Figure \ref{radii} the observed
radius--luminosity, radius--mass and the mass--luminosity relations are
compared with the theoretical prediction from the same Baraffe et al. (1998)
models.  We have estimated the masses of our stars using the empirical $K$ band
mass--luminosity calibration of Delfosse et al. (2000) which applies for $M_{K}
\ge 4.5$. For brighter luminosities we have used the empirical relation in $K$
band from Henry \& McCarthy (1993). The use of two different calibrations
---which are however fully consistent--- might be responsible for same small
offset, but the overall trend is well defined. We also convert our 2MASS
photometry into the CIT system (Carpenter 2001) before applying the
aforementioned empirical calibrations.  In panels (a) and (b) of Figure
\ref{radii} it is obvious that current models underestimate the radii of the M
dwarfs by $15-20$ percent, as already noticed by several other authors (see
Ribas 2006 for a review). Such a definite conclusion has been obtained using
double--lined eclipsing binaries. Our study reinforces the finding and confirms
its existence also to single field stars (Berger et al. 2006).  For the
mass--luminosity relation in Figure \ref{radii} (c), the disagreement between 
data
and theoretical models is less dramatic. In particular, going to masses below
$0.5\,M_{\odot}$ the agreement improve considerably as already noticed by other
authors (e.g. Delfosse et al. 2000). Since such a good agreement between the
data and the models is not present in the other panels of Figure \ref{radii}
neither in the temperature--luminosity plane of Figure \ref{bump}, it argues in
favour of a scenario in which the stars have larger radii and cooler effective
temperatures than predicted by models, but just in right proportion to barely
affect the luminosities.

In what follows we briefly discuss possible mechanisms responsible for the
radius increase which marks the transition from K to M dwarfs. An interesting
discussion on the disagreement between the predicted and measured radii of very
low mass stars from eclipsing binaries and interferometry can be found e.g. in
L\'opez-Morales (2007).  We also mention that another discontinuity at cooler
effective temperatures ($V-I_C \sim 2.7$), i.e. when the M dwarfs become fully
convective, is already known in literature (e.g. Hawley et al. 1996; Clemens et
al. 1998; Koen et al. 2002) and we do not discuss it here.

\subsection{Mixing-length}\label{mixle}

Very low mass stars are a very interesting place to test the input physics in
stellar models, since below $\sim 0.4 \, \rm{M}_{\odot}$ (depending on the
metallicity and the inclusion of magnetic fields in the models) stellar
interiors are expected to become fully convective. Their evolution is thus
practically insensitive to the mixing length parameters $\alpha_{\rm MLT}$ 
and the models thus are not subject to any adjustable parameter other than 
the helium abundance
(which is expected to be solar scaled). For this reason, very low mass models
do not need to be calibrated on the Sun.

The Baraffe et al. (1998) models are computed assuming a mixing length
$\alpha_{\rm MLT}=1$, quite different to values of $1.5-2$ which are typically
adopted for the Sun, and it is this which leads to the difference between the
models and data for the G and K dwarfs in Figure \ref{bump}. In fact, if a
solar calibrated model is used (Baraffe et al. 1998, continuous line), the
agreement for those stars becomes excellent. We have already extensively tested
theoretical models for G and K dwarfs in our previous paper (Casagrande et
al. 2007) and so we focus here on the M dwarfs.

A possible solution to the discontinuity in the HR diagram could be a rapid
decrease of the mixing-length as a function of stellar mass, although this 
would keep rather unaffected the lower part of the HR diagram in Figure 
\ref{bump}, where 
theoretical isochrones would still remain offset with respect to the observed 
stars. Since the
mixing-length describes the efficiency of the convection, any physical
mechanism inhibiting convection (like magnetic activity discussed in Section
\ref{magma}) can be phenomenologically mimicked by decreasing the mixing-length
(Chabrier, Gallardo \& Baraffe 2007).  Very interestingly, there are
indications of a possible dependence of the mixing-length with mass from
modelling the components of binaries (e.g. Lebreton, Fernandes 
\& Lejeune 2001; Yildiz et al. 2006). Just
how viable this solution is we regard as an open question.

\subsection{Magnetic activity}\label{magma}

The discrepancy between the predicted and observed radii in M dwarfs could be
related to the activity level of the stars (e.g. Torres \& Ribas 2002;
L\'opez-Morales \& Ribas 2005). It is known that active and inactive M dwarfs
define two different sequences in the luminosity-colour (Stauffer \& Hartmann
1986) and temperature-radius (Mullan \& MacDonald 2001) plane. Strong magnetic
fields are expected to inhibit convection, thus giving larger radii for a given
$\Teff$ or lower $\Teff$ for a given radius (Mullan \& MacDonald 2001).
Alternatively, it has also been suggested that the larger radii could simply be
an effect of flux conservation in a magnetic spot-covered stellar surface
(L\'opez-Morales \& Ribas 2005).

The stars with {\it Hipparcos} parallaxes plotted in Figure \ref{bump}, 
are expected to have a very low activity level because of the strict 
selection criteria used in Section \ref{sample}. This however does not exclude 
the possibility of a large and homogeneous spot-coverage since that would 
not necessarily produce any strong variability. To gauge further insight into 
the problem, we have checked that when the stars labeled as variable in Koen 
et al. (2002) are included in Figure \ref{bump}, they define the same trend 
shown by 
the stars of Section \ref{sample}. Notice though that the M dwarfs in Koen et 
al. (2002) were selected among the less variable in 
{\it Hipparcos}. Dedicated studies of active M dwarfs should still be done 
before reaching a more firm conclusion: at present, although magnetic 
activity can undoubtedly affect
stellar radii, we regard it as unlikely as being responsible for the main
sequence discontinuity observed at $4200-4300\,\rm{K}$. Concede the effect of
magnetic field certainly becomes more important descending the main sequence
and its inclusion is likely to be a relevant ingredient also for a proper
modelling of non-active M dwarfs.

\subsection{Opacity}

The models clearly have difficulty in reproducing the strong transition between
late K and early M type dwarfs, but the disagreement becomes even more
pronounced as one descends to the bottom of the main sequence. Similar
disagreement was already noticed by Baraffe et al. (1998) when comparing their
models with a sample of field stars, open and globular clusters with
metallicities similar to those covered in the present study. Since the
disagreement is much less pronounced in the comparison with metal poor globular
cluster M dwarfs (Baraffe et al. 1997), the disagreement at high metallicity
might be ascribed to missing opacity of some sort in the models.

\begin{figure*}
\begin{center}
\includegraphics[scale=0.55]{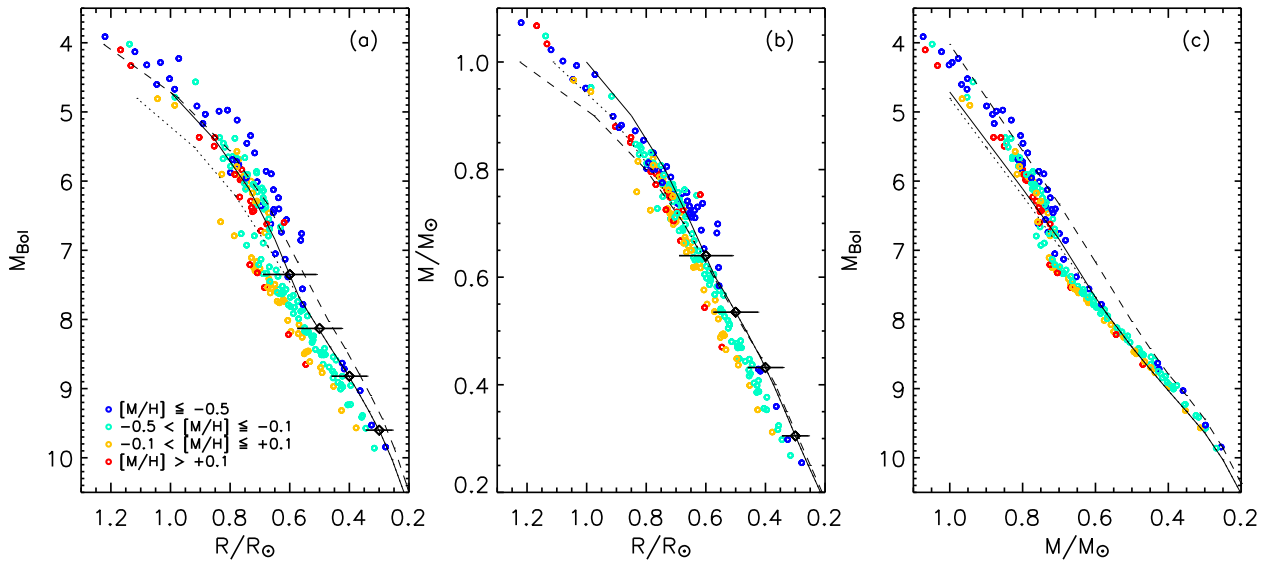}
\caption{Comparison with the Baraffe et al. (1998) models as in Figure
\ref{bump} but for the (a) radius--luminosity, (b) radius--mass and (c)
mass--luminosity relations. Error bars of $\pm 15$ percent in radius are shown
for $0.6, 0.5, 0.4, 0.3\,R_{\odot}$ for the solar metallicity model.}
\label{radii}
\end{center}
\end{figure*}

Figures \ref{bump} and \ref{radii} also indicate that the disagreement is more
marked for metal rich stars, again suggesting that missing opacity sources
could be a viable solution (Berger et al. 2006). It is very interesting that
the discontinuity occurs at a temperature when molecular formation
($\rm{H}_2\rm{O}$ and TiO) starts to be important, again suggesting that
opacity is a possible culprit.

\subsection{Three characters in search of an author}

We have briefly discussed three possible causes for the luminosity-temperature
discontinuity in the main sequence going from K to M dwarfs. The data in Figure
\ref{bump} and \ref{radii} suggest that the problem is more likely to be
related to molecular opacity, than being structural, but considerable further
work is needed to test those ideas. Simple steps forward to confirm or rule 
out possible explanations would be the analysis of late K and early M dwarfs' 
spectra as well as to run the MOITE for a large sample of (magnetically) 
active M dwarfs, to help
in searching for correlation between the radius discrepancy and activity level
or other physical parameters (e.g. L\'opez-Morales 2007). Of course, there may
be no a unique culprit for the radius discontinuity in M dwarfs, and only
advances in modelling both the structure and the atmosphere of these stars will
get things right.

\section{Conclusions}\label{conclu}

We have determined the temperature scale of M dwarfs, using stars with very
accurate multi--band photometry from optical to near-IR and the MOITE, a new
method which exploits the flux ratio in different bands as a sensitive
indicator of both effective temperatures and metallicities. Our proposed
temperature scale extends down to $\Teff \sim 2100-2200\,\rm{K}$ i.e. to the
L dwarf limit (e.g.  Leggett et al.  2002) and above $\sim 3000\,\rm{K}$ is
supported from interferometric angular diameters.  Our metallicities, which are
ultimately calibrated on Bonfils et al.'s (2005) metallicity scale, are also
found to be in very good agreement with the latest measurements from Woolf \&
Wallerstein (2005, 2006) and Bean et al. (2006a, 2006b), even if significant
differences in the various effective temperature scales still exist. Cool M
dwarfs with metallicities based on (hotter) \emph{Hipparcos} common
proper--motion companions also suggest our metallicities are reliable even
below $3000\,\rm{K}$, although further data are needed. Accurate
multi--band photometry for the coolest \emph{Hipparcos} common proper--motion
pairs would permit one to firmly extend the MOITE to the bottom of the main
sequence, thus opening this elusive area also to galactic chemical evolution
investigations. Exoplanets are found around M dwarfs, and a uniform metallicity
scale for their host stars will also be very useful.

The high quality of our data allows us to identify a striking feature which
marks the transition from K to M dwarfs, which appears to be due to an increase
in the radii of the early M dwarfs relative to late K dwarfs. We have compared
our sample of stars with theoretical isochrones for low mass stars and find
that such a feature is not predicted by the models, substantially confirming
the disagreement already noticed in the case of eclipsing binaries. Possible 
explanations including the effect of magnetic fields and molecular opacity 
have been discussed. 

This work also highlight the potentiality of high accuracy multi--band
photometry in determining fundamental stellar parameters and identifying fine
details in the HR diagram.  In particular, the MOITE will hugely benefit from
the existing infrared (2MASS, DENIS) and forthcoming optical surveys like
SkyMapper (Keller et al. 2007), Pan-Starrs (Kaiser et al. 2002) and LSST
(Claver et al. 2004) which will provide accurate and homogeneous multi--colour
and multi--epoch photometry for a large number of stars.

\section*{Acknowledgments}

LC acknowledges the Turku University Foundation and the Otto A. Malm Foundation
for financial support. This study was also funded by the Academy of Finland
(CF). We are indebted to C.\ Koen for providing the SAAO photometry and a 
careful reading on the first draft of the paper. 
We thank the PHOENIX team for making their models publicly available 
and P.\ Hauschildt and I.\ Brott for useful correspondence as well as 
M.\ Asplund for the same kindness. L.\ Portinari is also acknowledged for 
enlightening discussions. We are also indebted to B.\ Gustafsson for relevant 
comments and insight on many points of the paper. 
We thank an anonymous referee for many useful comments and suggestions
which have significantly improved the presentation of the paper.
The research has made use of the General Catalogue of Photometric Data
operated at the University of Lausanne and the SIMBAD data base, operated at
CDS, Strasbourg, France. The publication makes use of the data products from
the Two Micron All Sky Survey, which is a join project of the University of
Massachusetts and the Infrared Processing and Analysis Centre/California
Institute of Technology, funded by the National Aeronautics and Space
Administration and the National Science Foundation.

\appendix

\section[]{MOITE, technical details}

We use the Phoenix grid of synthetic spectra presented in Section 
\ref{phoenix} to bootstrap the MOITE. We assume $\log(g)=5.0$ throughout but 
we have tested that a change of $\pm 0.5$~dex in the assumed surface gravity 
has negligible effect on the results. 

For any given star in our sample, we first use the $\Teff : (R-I)_C$ 
calibration of Bessell (1991) to obtain an initial estimate of the effective 
temperature $T_\mathrm{eff,0}$. We then interpolate over our grid of Phoenix 
model atmosphere to compute the flux missing from our multi--band photometry 
and reconstruct the bolometric flux on the Earth. At each $n$--iteration a new
$T_\mathrm{eff,n}$ is obtained ---according to equation (\ref{eq_irfm}) or 
(\ref{eq_mfm})--- until $| T_\mathrm{eff,n}-T_\mathrm{eff,n-1} | < 1\,\rm{K}$ and 
the final solution is thus found. 
The rationale that motivates the choice between equation 
(\ref{eq_irfm}) or (\ref{eq_mfm}) 
will be discussed in the following of this Appendix. 
Notice that at each iteration the estimate of the monochromatic and 
bolometric fluxes also improve because of the 
improved effective temperature used to interpolate over the grid of model 
atmosphere. The quantity $R_{obs}$ thus is not exactly constant, but it 
depends ---quite weakly, indeed--- on the improved estimate of the effective 
temperature obtained at each step. 

To interpolate over the grid of model atmospheres, both
$T_\mathrm{eff,n}$ and [M/H] are needed. This is only possible for our 118 M
dwarfs with metallicities obtained from the Bonfils et al. (2005) calibration
(Section \ref{met_bon}). For the remaining stars [M/H] is estimated with the 
technique presented in Section \ref{met}.

The behaviour of $R_{theo}$ in different bands for 
various metallicities and effective temperatures is shown in Figure \ref{fr}.
In the infrared, $R_{theo}$ increases monotonically with increasing $\Teff$ 
above $\sim 4000\,\rm{K}$. If $R_{obs}$ is greater (smaller) 
than $R_{theo}$, at each iteration $T_\mathrm{eff,n}$ increases (decreases) 
until it converges to its limiting value. In fact, let us consider the 
following case: 
\begin{equation}
R_{obs} > R_{theo}
\end{equation}
which implies
\begin{equation}
\frac{\mathcal{F}_{Bol}\textrm{(Earth)}_{(n-1)}}
{\mathcal{F}_{\lambda}\textrm{(Earth)}_{(n-1)}} > \frac{\sigma T_{\mathrm{eff},n-1}^{4}}{\mathcal{F}_{\lambda}\textrm{(model)}_{(n-1)}}
\end{equation}
and rearranging to highlight the result
\begin{displaymath}
T_\mathrm{eff,n} = \left( \frac{\mathcal{F}_{\lambda}\textrm{(model)}_{(n-1)} 
\mathcal{F}_{Bol}\textrm{(Earth)}_{(n-1)}}
{\sigma \mathcal{F}_{\lambda}\textrm{(Earth)}_{(n-1)}}\right)^{\frac{1}{4}}
\end{displaymath}
\begin{equation}
\phantom{T_\mathrm{eff,n} =} > T_\mathrm{eff,n-1}.
\end{equation}
The case $R_{obs} < R_{theo}$ can be similarly proven to give 
$T_\mathrm{eff,n} < T_\mathrm{eff,n-1}$.

\begin{figure}
\begin{center}
\includegraphics[scale=0.55]{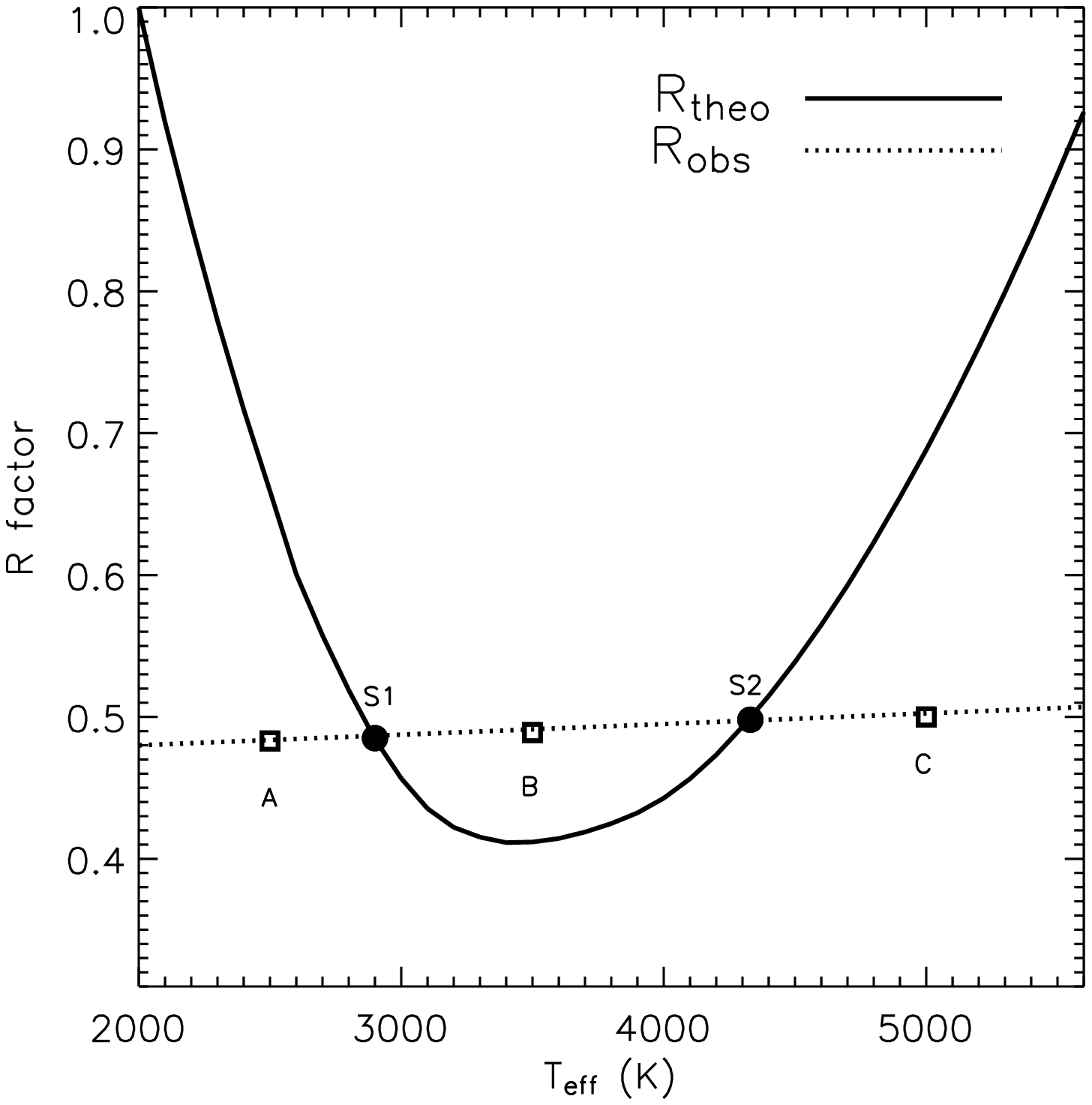}
\caption{Schematic representation of the degeneracy in the temperature solution
in a given band using the IRFM for cool stars. Arbitrary units are used on the
$y$-axis. At low temperature, $R_{obs}$ intersects $R_{theo}$ twice, producing
two solutions: S1 and S2 (filled circles). If the IRFM starts from A, then
$R_{obs} < R_{theo}$ and at each step the new temperature estimate decreases,
and diverges away from the desired solution, S1. In case B, $R_{obs} >
R_{theo}$ and at each step the new temperature estimate increases until it
reaches the hotter solution S2 and terminates. In neither case is the cooler
solution S1 found. In case C, $R_{obs} < R_{theo}$ and the IRFM also converges
onto solution S2. The use of equation (\ref{eq_irfm}) in the IRFM thus finds 
only
the hotter of two temperature solutions and can not be used for very cool
stars. A technique which finds the appropriate solution over many bands has
been developed in this paper, and the technical details are discussed here.}
\label{sketch}
\end{center}
\end{figure}

The technique thus converges quickly above $\Teff$ about $4000\,\rm{K}$, even 
with quite poor initial estimates of the effective temperature, as it can be 
more readily understood by looking at the sketch of Figure \ref{sketch}.
Going to cooler effective temperatures, a given $R_{obs}$ intersects 
$R_{theo}$ twice, i.e. there are two possible iterative solutions. In the case 
of a very cool star (say below $3500\,\rm{K}$ in the example of Figure 
\ref{sketch}), if $R_{obs}$ is greater than $R_{theo}$, at each iteration
$T_\mathrm{eff,n}$ always increases to the solution with highest effective
temperature. Similarly, if $R_{obs}$ is smaller than $R_{theo}$ then at each
iteration $T_\mathrm{eff,n}$ continues decreasing without reaching a
solution. Thus, equation (\ref{eq_irfm}) cannot be used for the cooler stars
since it only finds one temperature (the hotter one) or none at all. An
alternative approach to overcome this limitation would be to sample the entire
$R_{theo}$ space, find the two effective temperatures that minimize $| R_{theo}
- R_{obs} |$ and choose the proper solution. Other then being more
computationally demanding, at cool temperatures the two minima are quite
shallow in the infrared since $R_{theo}$ inverses smoothly and it might not be
obvious which one of the two solutions must be chosen.  
On the contrary, the use of the flux products when $R_{theo}$ increases with 
decreasing $\Teff$ allows to converge at cool temperatures via equation 
(\ref{eq_mfm}).

It is clear that in any given band, when $R_{theo}$ increases with increasing
effective temperature equation (\ref{eq_irfm}) must be used, whereas when
$R_{theo}$ increases with decreasing effective temperature equation
(\ref{eq_mfm}) must be used.  Only when $R_{theo}$ inverts any dependence on
the effective temperature is lost. Fortunately, using multi--band photometry,
at any given $\Teff$ is always possible to find one or more bands for which
$R_{theo}$ has a well defined behaviour, i.e. is either monotonically
increasing or decreasing, discarding bands for which the dependence is
practically flat.

Above $4000\,\rm{K}$ we use equation (\ref{eq_irfm}) to estimate 
$T_\mathrm{eff,n}$
from $J, H, K_S$ photometry identically to Casagrande et al. (2006).  Note that
the choice between equation (\ref{eq_irfm}) or (\ref{eq_mfm}) is important to
correctly converge to $T_\mathrm{eff,n}$. However, when the final solution is
found, equation (\ref{eq_irfm}) or (\ref{eq_mfm}) returns effective
temperatures that agree within $\sim 5\,\rm{K}$ in any given band. For this 
reason,
once a solution is found it is possible to have an estimate of the effective
temperature from all the other $\xi$ bands ($T_{\xi}$). The use of both 
optical and infrared colours is crucial for estimating the metallicities, as 
we explain in Section \ref{met}. 
In the optical we
estimate $T_{\xi}$ from $V, R_C, I_C$ photometry; we did not use the $U$ and
$B$ bands since these colours are not available for all the stars and their
theoretical modeling is also more uncertain.

For $2500 < \Teff < 4000\,\rm{K}$ we use equation (\ref{eq_irfm}) to estimate
$T_\mathrm{eff,n}$ from $J$ band and equation (\ref{eq_mfm}) to estimate
$T_\mathrm{eff,n}$ from $V$ and $R_C$ band. Below $2500\,\rm{K}$ we also use 
equation
(\ref{eq_mfm}) to estimate $T_\mathrm{eff,n}$ from $I_C$.  We then average the
results obtained in these bands for the next iteration.  Again, when a solution
is found we compute the effective temperatures $T_{\xi}$ predicted by all the
colours (with the exception of $U$ and $B$ bands, as we already said).  For the
most metal poor stars, $I_C$ and $J$ bands flatten out at very cool
temperatures. As suggested by Figure \ref{fr} we have also implemented a more 
refined approach, to ensure that we always use as many bands as possible for 
which $R_{theo}$ is expected to monotonically increase or decrease. Our code, 
written in IDL, is available upon request.

It is important to note that our technique assigns equal weight to each of the
bands in the converge to an effective temperature. It might plausibly be
improved by identifying the bands which are more sensitive to effective
temperature and those to metallicity in determining these parameters. In this
sense, the IRFM can be regarded as a more elegant technique to determine
$\Teff$, since it works in the Rayleigh--Jeans part of a spectrum and is not
much affected by the metallicity. However, below $\sim4000\,\rm{K}$ practically
all bands start to show considerable dependence on the metallicity, the only
exception being $J$ band (Figure \ref{fr}). The IRFM is not quite metallicity
independent in any case, since the reconstruction of the bolometric flux from
multi--band photometry still depends on the metallicity used to interpolate in
the grid of model atmospheres. We have looked at two scenarios in which MOITE
may need to perform, such as only IR or optical photometry being available. 

Firstly, we have checked whether any major difference arises by using only $J$
band to determine effective temperatures. The behaviour of the flux ratio in 
such band is in fact expected to be quite unique, with very little metallicity 
dependence and always increasing as function of $\Teff$ (Figure \ref{fr}). 
We did not find any considerable improvement, but only a mean temperature 
difference of $48 \pm 57 \,\rm{K}$, which we think it stems from the zero-point 
uncertainties in the $J$ band absolute calibration (uncertainty which instead 
average out using many bands). In addition, relying on one band means the 
technique is much more exposed to the quality of the photometry in that band.

Secondly, we have studied how the convergence in $\Teff$ is affected using only
$BVR_C$ colours. According to Figure \ref{fr} in the temperature range expected
for our stars, it should be possible to use optical colours only. We regard the
metallicity as a fixed known parameter and we compute the temperature
difference with respect to $\Teff$ obtained using both optical and infrared
colours. We have tested also the difference when infrared colours are still
used to recover the bolometric flux but not for converging in $\Teff$ and when
the infrared colours are not used at all. Summarizing, the mean difference is
of order $15 \pm 35\,\rm{K}$. This result is very
reassuring and also makes the technique promising to be used for M dwarfs for
which only optical colours are available. 

Therefore, at present, the use of all or only of some optical and infrared
bands seems to return reliable and consistent results.  We plan to further test
our findings in forthcoming studies, by addressing specifically the sensitivity
of different spectral bands to effective temperature and metallicity and
eventually refine the technique presented here.

\section[]{The Vega and Sirius absolute calibration}

In this work we have updated the absolute calibration of Vega in the optical 
by adopting the new reference spectra of Bohlin (2007) rather than that of 
Bohlin \& Gilliland (2004) and which is expected to be accurate within 
$\sim 1$ percent in the range $3200-10000$ \AA. 
In the infrared the absolute calibration of Vega is kept the same as 
in Casagrande et al. (2006), which is based on Cohen et al. (2003).
In terms of zero-points, the updated fluxes of Bohlin (2007) corresponds to 
changes of few millimag and affect the derived $\Teff$ by $10\,\rm{K}$, thus 
confirming the results obtained in Casagrande et al. (2006).

For some of our stars, we have also $U$ photometry (Section \ref{jc}).
For Vega we adopt $U=0.02$ and the same magnitudes as in Casagrande et al. 
(2006) for the other bands (i.e. $BV(RI)_CJHK_S$). $U$ filters has proven 
rather difficult to standardize (e.g. Bessell 1986, 1990b) and adjustment 
of the $U$ zero-points for different temperature ranges has also been discussed 
(Bessell, Castelli \& Plez 1998). $U$, $U-B$ and $B-V$ colours are computed 
according to the prescription in Bessell (1990b).

The Vega zero-points and absolute calibration thus seem now firmly established 
in the optical (Bohlin \& Gilliland 2004; Bohlin 2007), but some 
uncertainties (that however do not exceed few percents) still remain in the 
infrared due to its pole-on and rapidly rotating nature.  
The IRFM and MOITE temperature scales are intimately related to the adopted 
infrared zero-points and absolute calibration. The possibility of basing our 
technique on a different photometric system and standard star is a valuable 
sanity check to the proposed temperature scale.

The SAAO $JHK$ photometric system was established by Glass (1974) and its
accuracy and zero-points refined and improved over the years by Carter (1990)
and Carter \& Meadows (1995). Since Vega is unobservable in the Southern
hemisphere, the zero-points of the SAAO $JHK$ photometric system are based on
25 main sequence stars ranging from spectral type B1 to A7 (Carter, 1990).
Sirius is often chosen as an complementary or alternative standard to Vega
(e.g. Cohen et al. 1992). Its observed magnitudes and colours in the SAAO $JHK$
photometric system are given in Table \ref{sirius}.
\begin{table}
\centering
\caption{Observed magnitudes for Sirius in the SAAO $JHK$ system.}
\label{sirius}
\begin{tabular}{cccr}
\hline
   $J$   &  $H$    &  $K$     & Ref.\\
\hline
 $-1.387$  & $-1.378$  & $-1.369$   & Bessell et al. 1998 \\
\hline
\end{tabular}
\end{table}

Since no absolute flux measurements are available for Sirius, Cohen et
al. (1992) decided to absolutely calibrate a Kurucz (1991) Sirius model with
respect to Vega, by using observed magnitude difference between Vega and Sirius
in different near and mid infrared bands. Their resulting angular diameter for
Sirius was $\theta = 6.04$~mas, $0.9 \, \sigma$ larger than the direct
measurement (corrected for limb-darkening) $\theta = 5.89 \pm 0.16$~mas by
Hanbury Brown, Davis \& Allen (1974).  
Recently, new interferometric measurements have
become available for Sirius. Davis \& Tango (1986) obtained $\theta = 5.93 \pm
0.08$ (when updated limb-darkening coefficients are used, see Kervella et
al. 2003), while Mozurkewich et al. (2003) found $\theta = 5.993 \pm
0.108$. All these direct measurements however were obtained at optical
wavelength, where the limb-darkening corrections are larger and more difficult
to assess. Recently Kervella et al. (2003) have observed Sirius in the near
infrared, where the limb-darkening corrections are much smaller, obtaining
$\theta = 6.039 \pm 0.019$~mas, in superb agreement with spectrophotometric
value of Cohen et al. (1992).

We absolutely calibrate Sirius by scaling its latest Kurucz (2003) synthetic
spectrum with the angular diameter measurement of Kervella et al. (2003) and
therefore independently of any consideration about Vega.  The corresponding
effective wavelength and absolute calibration in the SAAO $JHK$ filters are
reported in Table \ref{abssi}.  The error in the angular diameter given by
Kervella et al. (2003) implies an uncertainty of only 0.6 percent in
monochromatic absolute fluxes. We adopt a more conservative approach, by taking
the standard deviation from all the aforementioned interferometric 
measurements: these give an uncertainty of $0.066$~mas that translates into an 
uncertainty
of circa 2 percent in fluxes, in good agreement with the global uncertainty of
1.46 percent estimated by Cohen et al. (1992). Also, the fact the dominant H
opacity in A stars is expected to be well understood gives confidence on the
adoption of a synthetic spectrum.
\begin{table*}
\centering
\caption{Absolute calibration and effective wavelength of the ground-based 
SAAO $JHK$ photometry of Sirius. Quantities tabulated correspond to the 
definition of the zero magnitude in each filter.}\label{abssi}
\label{sirius_abs}
\begin{tabular}{cccc}
\hline
Band & $\lambda_\mathrm{eff}$ & Monochromatic Absolute Flux & Uncertainty\\
\hline
 & \AA & erg cm$^{-2}$ s$^{-1}$\AA$^{-1}$ & erg cm$^{-2}$ s$^{-1}$\AA$^{-1}$\\
\hline
$J$  &	12044  &  1.176e$-$09  &  2.570e$-$11\\
$H$  &	16282  &  4.079e$-$10  &  8.916e$-$12\\
$K$  &	22004  &  1.367e$-$10  &  2.988e$-$12\\
\hline
\end{tabular}
\begin{minipage}{1\textwidth}
The Kurucz model adopted for Sirius has $\Teff=9850\,\rm{K}$, $\log(g)=4.3$,
$\textrm{[M/H]}=+0.4$ and microturbolent velocity $\xi = 0 \;\textrm{km} \; 
\textrm{s}^{-1}$. The same formalism adopted in Casagrande et al. (2006) is
used. Notice that the SAAO $JHK$ photometer is equipped with a InSb detector 
and therefore in generating fluxes from model atmosphere energy-integration 
is the most appropriate. 
\end{minipage}
\end{table*}

We have run the MOITE for stars with SAAO $JHK$ photometry 
(Section \ref{nir}), adopting the zero-points and absolute 
calibration of Vega in $UBV(RI)_C$ and of Sirius in $JHK$ 
(Table \ref{sirius} and \ref{sirius_abs}). The difference with respect to 
the use of the 2MASS $JHK_S$ photometry and absolute calibration (Cohen et 
al. 2003) is negligible, thus confirming the adequacy of the absolute 
calibration adopted in this work and in Casagrande et al.\ (2006).
The mean difference in  $\Teff$ is $9 \pm 3\,\rm{K}$ ($\sigma=25\,\rm{K}$), 
and in both bolometric luminosity and angular diameters is well below 1 percent.
The results provided by the adoption of the absolute calibration 
of Vega or Sirius are therefore identical within the errors.

\end{document}